\documentclass[twocolumn,showpacs,preprintnumbers,amsmath,amssymb]{revtex4}

\usepackage{epsf}
\usepackage{rotating}

\usepackage{graphicx,epsfig}
\usepackage{dcolumn}
\usepackage{bm}
\def\cA{{\cal A}}
\def\cD{{\cal D}}
\def\cN{{\cal N}}
\def\cL{{\cal L}}
\def\sD{{\mbox{\sf D}}}
\def\sN{{\mbox{\sf N}}}
\newcommand{\vgap}{\\ \vspace*{2.3mm}}

\begin{document}

\preprint{}

\title[Inference and resource optimization on sparse graphs]
{Inference and optimization of real edges on sparse graphs -
a statistical physics perspective}

\author{K.~Y.~Michael Wong$^{1}$ and David Saad$^{2}$}
\affiliation{$^{1}$ Department of Physics, The Hong Kong University
of Science and Technology, Hong Kong, China\\
$^{2}$ Neural Computing Research Group, Aston University,
Birmingham B4 7ET, UK}

\date{\today}

\begin{abstract}
Inference and optimization of real-value edge variables in sparse
graphs are studied using the Bethe approximation and replica
method of statistical physics. Equilibrium states of general
energy functions involving a large set of \emph{real}
edge-variables that interact at the network nodes are obtained in
various cases. When applied to the representative problem of
network resource allocation, efficient distributed algorithms are
also devised. Scaling properties with respect to the network
connectivity and the resource availability are found, and links to
probabilistic Bayesian approximation methods are established.
Different cost measures are considered and algorithmic solutions
in the various cases are devised and examined numerically.
Simulation results are in full agreement with the theory.

\end{abstract}

\maketitle


\section{Introduction}
\label{sec:Introduction}

The links between statistical physics models and a variety of
inference and optimization problems have been significantly
strengthened over the last decade~\cite{Nishimori_book}. Two
aspects of these links have been exploited. Macroscopically, using
the statistical physics framework, one describes typical
properties of the problem and provides valuable insight into its
generic characteristics. Microscopically, established techniques
of statistical physics such as the cavity method have been used
for devising efficient inference algorithms, some of which have been
independently discovered and used in other research
communities~\cite{os,mackaybook,yedidia,mezard}.

Most studies so far, both within and outside the statistical physics
community, have focused on cases of discrete variables.
Among the recently successful examples
using methods of statistics-based mechanics, one can mention hard
computational problems~\cite{MZreview}
and error-correcting codes~\cite{KSreview}.
Statistical mechanical approaches to learning of discrete variables
have also been considered on tree structures~\cite{saul94}.

On the other hand, networks of continuous variables
were much less explored.
One of the main reasons for this limited activity
is the difficulty in applying message
passing approximation algorithms~\cite{os,mackaybook} in this
case, as the discrete messages passed between variables become
{\em functions} of real variables. Applied message passing for systems
of real variables typically relies on modelling the functions
using a reduced number of parameters~\cite{Lauritzen}.

In the statistical physics community there have been recent
attempts to simplify the messages for continuous variables. 
For example, a step forward was made in Ref.~\cite{skantzos} 
to parametrize the messages using eigenfunction decomposition
for special cases.
Furthermore, the continuous variables treated by these methods
are {\it localized} on nodes,
whereas many interesting problems,
such as the resource allocation problem presented here
(and partially in~\cite{us_nips,us_prerc})
involves real variables defined on links {\it between} nodes.

In this paper we study a system with real variables that can be mapped
onto a sparse graph and suggest an efficient message-passing
approximation method for inference and optimization. We first
formulate the problem at a general temperature; the message-passing
algorithm we present here as well as the related analysis are
primarily {\em general} inference algorithms. In this
paper, however, we are particularly interested in the optimal, zero
temperature, solution that reduces the task to an optimization
problem.

Global optimization techniques, such as linear or quadratic
programming~\cite{bertsekas} can successfully solve many of these
problems. However, message-passing approaches have the potential
to solve global optimization problems via {\em local} updates,
thereby reducing the growth in computational complexity from cubic
to linear with the system size. An even more important practical
advantage is its distributive nature that is particularly suitable
for distributive computation in large or evolving networks and
does not require a global optimizer.

We focus on a prototype for optimization,
and use the example of resource allocation
as a vehicle to demonstrate the potential of our method,
both for gaining insights into the main properties of the
system and as an efficient optimization algorithm.
Our method is efficient
since the messages consist of only the first and second derivatives
of the vertex free energies derived from our analysis.
The key to the successful simplification,
not needed for the simpler case of discrete variables,
is that the messages passed to a target node
are accompanied by information-provision messages from the target node,
to first determine the working point
at which the derivatives should be calculated.

The problem of resource allocation is a well known network
problem in the areas of computer science and operations
management~\cite{resourceallocation,resourceallocation2}.
The problem itself is quite general and is applicable to typical
situations where a large number of nodes are required to balance
loads/resources, such as reducing internet traffic congestion and
streamlining network flow of commodities~\cite{Shenker,om}. In
computer science, many practical algorithmic solutions have been
proposed to distribute computational load between computers
connected in a network. They usually are heuristic and focus on
practical aspects (e.g., communication protocols). The problem we
are addressing here is more generic and, in the context of
computer networks, is represented by nodes of some computational
power that should carry out tasks; sub-tasks are then moved around
such that all demands will be satisfied while the migration of
(sub-)tasks is minimized.

In section~\ref{sec:model} we will introduce the general model,
followed by a replica-based analysis in section~\ref{sec:replica} and
subsequently by a Bethe approximation-based analysis in
section~\ref{sec:bethe}. A message passing algorithm for the problem
of resource allocation will be presented in
section~\ref{sec:algorithm} followed by the derivation of scaling laws
in the limit of high connectivity in
section~\ref{sec:asymptotic}. Numerical results for
sections~\ref{sec:replica}-\ref{sec:asymptotic} will be presented in
section~\ref{sec:numerical}.  We will then extend the model to the
case of general cost functions in section~\ref{sec:generalcost},
highlighting strengths and weaknesses of our approach. We will
conclude the presentation with a summary and point to future
research directions.

\section{The model}
\label{sec:model}

The problem we are addressing here is a generic version of
resource allocation and serves as an example of a sparsely
connected system of real variables that should be optimized with
respect to some general cost. It is represented by nodes of some
computational power that should carry out tasks. Both
computational powers and tasks will be chosen at random from some
arbitrary distribution. The nodes are located on a randomly chosen
sparse network of some connectivity. The goal is to migrate tasks
on the network such that demands will be satisfied while the
migration of (sub-)tasks is minimized. We focus here on the
satisfiable case where the total computing power is greater than
the demand, and where the number of nodes involved is very large.

The sparse network considered has $N$ nodes, labelled
$i\!=\!1,\dots,N$. Each node $i$ is randomly connected to $c$
other nodes. The connectivity matrix is given by 
$\cA_{ij}=\cA_{ji}=1, 0$
for connected and unconnected node pairs respectively. A link
variable $y_{ij}$ is defined on each connected link from $j$ to
$i$. We focus on the case of intensive connectivity $c\sim O(1)\ll
N$; and restrict the problem to the fixed connectivity case
although both the analysis and the algorithm devised on its basis
can handle a general connectivity profile.

We consider a general energy function (cost)
\[ E=\sum_{(ij)}\cA_{ij}\phi(y_{ij})+\sum_i
\psi(\Lambda_i,\{y_{ij}|\cA_{ij}=1\})
 \ , \] where the summation $(ij)$ is made over all node pairs,
and $\Lambda_i$ is a quenched variable defined on node
$i$. In the context of probabilistic inference, $y_{ij}$ may represent
the coupling between observables in nodes $j$ and $i$, $\phi(y_{ij})$
may correspond to the logarithm of the prior distribution of $y_{ij}$,
and $\psi(\Lambda_i,\{y_{ij}|\cA_{ij}=1\})$ the logarithm of the
likelihood of the observables $\Lambda_i$.
Since the cost is independent of the direction of the currents 
in many applications, we focus on the case 
that $\phi(y)$ is a general even function of $y$.
In the context of resource allocation,
$y_{ij}\equiv-y_{ji}$ may represent the current from node $j$ to $i$,
$\phi(y_{ij})$ may correspond to the transportation cost, and
$\psi(\Lambda_i,\{y_{ij}|\cA_{ij}=1\})$ the performance cost of the
allocation task on node $i$,
dependent on the node capacity $\Lambda_i$;
the capacity of a node is defined as
its computational capability minus its computational demand,
and is randomly drawn from a distribution $\rho(\Lambda_i)$.

\section{Replica analysis}
\label{sec:replica}

To make the analysis more concrete and strengthen
the link to the resource allocation problem, 
we keep the term $\phi(y_{ij})$ general and, 
aiming to satisfy the capacity constraints, 
set $\psi(\Lambda_i,\{y_{ij}|\cA_{ij}=1\}) =\ln[\Theta(-\sum_j
\cA_{ij}y_{ij}-\Lambda_i)+\epsilon]$, 
where $\epsilon\to 0$ and $\Theta$ is the step function.
This reduces the problem to the load balancing task
of minimizing the energy function (cost)
$E=\sum_{(ij)}\cA_{ij}\phi(y_{ij})$,
subject to the constraints on the resources of nodes $i$,
\begin{equation}
        \sum_j \cA_{ij}y_{ij}+\Lambda_i\ge 0 \ ,
\label{constr}
\end{equation}

An alternative formulation is to consider the dual of the original
optimization problem. Introducing Lagrange multipliers, the function
to be minimized becomes
\begin{equation}
        L\!=\!\sum_{(ij)}\cA_{ij} \phi(y_{ij})
        +\sum_i\mu_i \left(\sum_j \cA_{ij}y_{ij}+\Lambda_i \right)\ .
\label{lagr}
\end{equation}
Optimizing $L$ with respect to $y_{ij}$, one obtains
\begin{equation}
        y_{ij}\!=\![\phi']^{-1}(\mu_j-\mu_i),
\label{yij}
\end{equation}
where $\mu_i$ is referred to as the {\it chemical potential} of node $i$,
and $\phi'$ is the derivative of $\phi$ with respect to its argument.
This can be interpreted as the current being driven
by the potential difference.

Since the probability of finding loops of finite lengths is
vanishing in large sparse networks,
the structure of a sparse network is locally a tree.
Thus, given a configuration of currents $\{y_{ij}\}$,
one can set the {\it current potential} $\nu_i$ of a node
to an arbitrary value,
and assign $\nu_j$ of its neighbors according to
$\nu_j=\nu_i+y_{ij}$.
Repeating this assignment process to next nearest neighbors and so on,
the current potentials of all nodes in the tree can be determined.
Hence, the current potentials can be considered
as valid independent variables as the current variables used originally.
This implies that we can consider
the optimization problem in the space
of the current potentials. Since the energy function is invariant
under the addition of an arbitrary global constant to the current
potentials of all nodes, we introduce an extra regularization term
$\epsilon\sum_i\mu_i^2/2$ to break the translational symmetry,
where $\epsilon \rightarrow 0$.
(Note that the current potentials $\nu$
are different from the chemical potentials $\mu$,
which are the Lagrange multipliers
of the dual formulation in Eq.~(\ref{lagr}).
Only for the quadratic cost $\phi(y)=y^2/2$
can the current be expressed
in terms of the difference in chemical potentials.
Even in this case, the two potentials may differ
by a non-vanishing constant since the resource constraints
in Eq.~(\ref{constr}) imply
that the maximum of the Lagrange multipliers is 0, whereas 
the current potentials minimize $\epsilon\sum_i\nu_i^2/2$ and are
unlikely to have a maximum value of 0.)
The corresponding partition function is
\begin{eqnarray}
\label{eq:Zchem}
        &&Z=\prod\limits_i \int
        d\nu _i \prod\limits_i \Theta \left[ \sum\limits_j \cA_{ij}
        (\nu_j -\nu _i ) +\Lambda _i \right]
        \nonumber\\
        &&\times\exp \left[ -\beta
        \sum\limits_{(ij)} \cA_{ij} \phi(\nu _j -\nu _i )
        -\frac{\beta\epsilon}{2}\sum_i\nu_i^2 \right] .
\end{eqnarray}

The replicated partition function~\cite{Nishimori_book}, at a
temperature $T\equiv\beta^{-1}$, averaged over all network
configurations of connectivity $c$ and capacity distributions
$\rho(\Lambda_i)$, is given by
\begin{eqnarray}
        &&\langle Z^n\rangle_{\cA,\Lambda}
        \!=\!\frac{1}{\cal N} \sum_{ \cA_{ij}=0,1}
        \prod_i\Biggl\{\delta\left(\sum_j  \cA_{ij}-c\right)
        \int d\Lambda_i
        \nonumber\\
        &&\times\rho(\Lambda_i)
        \prod_{\alpha=1}^n\biggl[\int d\nu_i^\alpha
        \Theta\biggl(\sum_j  \cA_{ij}(\nu_j^\alpha-\nu_i^\alpha)
        +\Lambda_i\biggr)\biggr]\Biggr\}
        \nonumber\\
        &&\times\exp\left(-\beta\sum_{(ij)\alpha}
        \cA_{ij}\phi(\nu_j^\alpha-\nu_i^\alpha)
        -\frac{\beta\epsilon}{2}\sum_{i\alpha}(\nu_i^\alpha)^2\right).
\label{eq:replicatedpart}
\end{eqnarray}
Here ${\cal N}\!=\!\sum_{ \cA_{ij}=0,1}\prod_i\delta(\sum_j
\cA_{ij}\!-\!c)$ is the total number of graphs with connectivity
$c$. This can be easily shown to be~\cite{wong87} ${\cal N}=\exp
\left\{ {N\left[ {-(c/2)+(c/2)\ln (cN)-\ln
c!}\right]} \right\}$.

The interaction coupling current potentials of different nodes makes
it difficult to decouple them in order to define macroscopic order
parameters.
Nevertheless, additional expansions detailed in Appendix A
also show that it is possible to dientangle neighboring node indices.
(This justifies the formulation of the optimization
in the space of the current potentials $\{\nu_i\}$
rather than that of the currents $\{y_{ij}\}$.)
This leads to the following definition of the order parameters
\begin{equation}
\label{eq:Qdef}
        Q_{\mathbf{{r}},\mathbf{{s}}}
        =\frac{1}{\sqrt {cN} }\sum\limits_i
        {z_i \exp \left({\sum\limits_\alpha
        {i\hat {\lambda }_i^\alpha \nu _i^\alpha } }\right)
        \prod\limits_\alpha {(-i\hat {\lambda }_i^\alpha )^{r_\alpha }
        (\nu_i^\alpha )^{s_\alpha }} },
\end{equation}
and its conjugate $\hat Q_{{\mathbf r},{\mathbf s}}$.  Following the
analysis of~\cite{wong87} and averaging over the connectivity tensor
$\cA$ one finds
\begin{eqnarray}
        &&\langle Z^n\rangle_{\cA,\Lambda}=\exp N\Biggl\{\frac{c}{2}
        -c\sum_{{\mathbf r},{\mathbf s}}
        \hat Q_{{\mathbf r},{\mathbf s}}Q_{{\mathbf r},{\mathbf s}}
        \nonumber\\
        &&+\ln\int d\Lambda\rho(\Lambda)\prod_\alpha\left(
        \int d\nu_\alpha\int_{-\Lambda}^\infty d\lambda_\alpha
        \int\frac{d\hat\lambda_\alpha}{2\pi}\right)
        \nonumber\\
        &&\times\exp\left[\sum_\alpha\left(i\hat\lambda_\alpha
        (\lambda_\alpha+c\nu_\alpha)
        -\frac{\beta\epsilon}{2}(\nu_\alpha)^2\right)\right]
        X^c\Biggr\},
\label{zn}
\end{eqnarray}
where
\begin{eqnarray}
\label{xx}
        &&X=\sum_{{\mathbf r},{\mathbf s}}
        \hat Q_{{\mathbf r},{\mathbf s}}
        \prod_\alpha(-i\hat\lambda_\alpha)^{r_\alpha}(\nu_\alpha)^{s_\alpha}
        \\
        &&+\frac{1}{2}\sum_{{\mathbf r},{\mathbf s}}
        Q_{{\mathbf r},{\mathbf s}}
        \prod_\alpha \frac{(\nu_\alpha)^{r_\alpha}}{r_\alpha!s_\alpha!}
        \left(-i\hat\lambda_\alpha-\frac{d}{dy}\right)^{s_\alpha}
        e^{-\beta\phi(y)}\Biggr|_{y=\nu_\alpha}.\nonumber
\end{eqnarray}
The somewhat unusual indices of the order parameters $Q_{{\mathbf
r},{\mathbf s}}$ and $\hat Q_{{\mathbf r},{\mathbf s}}$, the
vectors ${\mathbf r}$ and ${\mathbf s}$, represent $n$-component
vectors $(r_1,\dots,r_n)$ and $(s_1,\dots,s_n)$ respectively. This
is a result of the specific interaction considered which entangles
nodes of different indices.  The order parameters $Q_{{\mathbf
r},{\mathbf s}}$ and $\hat Q_{{\mathbf r},{\mathbf s}}$ are given
by the extremum condition of Eq.~(\ref{zn}), i.e., via a set of
saddle point equations with respect to the order parameters.
To facilitate the solution,
we introduce the generating function of $P_{\mathbf s}({\mathbf z})$
and its inversion,
\begin{eqnarray}
        &&P_{\mathbf s}({\mathbf z})
        \!=\!\sum_{\mathbf r}
        Q_{{\mathbf r},{\mathbf s}}
        \prod_\alpha\frac{(z_\alpha)^{r_\alpha}}{r_\alpha!},
        \nonumber\\
        &&Q_{{\mathbf r},{\mathbf s}}
        =\prod_\alpha\left(\frac{\partial}{\partial z_\alpha}\right)
        ^{r_\alpha}P_{\mathbf s}(\mathbf z)\Biggr|_{\mathbf z = 0}\ .
\label{genf}
\end{eqnarray}
Eliminating $\hat Q_{{\mathbf r},{\mathbf s}}$, and substituting
the saddle point equation of $Q_{{\mathbf r},{\mathbf s}}$ into
$P_{\mathbf s}({\mathbf z})$ in Eq.~(\ref{genf}), one finds
the recursion relation
\begin{eqnarray}
        &&P_{\mathbf s}({\mathbf z})=\frac{1}{\sD_P}
        \int {d\Lambda \rho (\Lambda )} \prod\limits_\alpha \left[
        \int {d\nu _\alpha } \int_{-\Lambda }^\infty {d\lambda _\alpha }
        \int \frac{d\hat {\lambda }_\alpha }{2\pi }\right.
        \nonumber\\
        &&\left.\times\exp \left( i\hat {\lambda}_\alpha
        (\lambda _\alpha +c\nu _\alpha -z_\alpha )
        -\frac{\beta \varepsilon}{2}(\nu _\alpha)^2 \right) \right]
        \sum\limits_{\mathbf{s}_k} \prod\limits_{k=1}^{c-1}
        \nonumber \\
        &&\times P_{\mathbf{s}_k}(\mathbf{\nu})
        \prod\limits_{k\alpha } \frac{1}{s_k^\alpha !}
        \left.\left(-i\hat\lambda_\alpha -\frac{d}{dy}\right)^{s_k^\alpha }
        e^{-\beta \phi(y)} \right|_{y=\nu_\alpha}
        \nonumber\\
        &&\times\prod\limits_\alpha(\nu _\alpha)^{s_\alpha},
\label{eq:P_sDef}
\end{eqnarray}
where $\sD_P$ is a constant given in Eq.~(\ref{eq:Dpdef}).
Note that $P_{\mathbf s}({\mathbf z})$ is 
expressed in terms of
$c\!-\!1$ functions $P_{\mathbf s_k}(\mathbf\nu)$
($k\!=\!1,..,c\!-\!1$), integrated over ${\mathbf\nu}$ and summed
over ${\mathbf s}_k$. This structure is typical of the Bethe
lattice description of networks of connectivity $c$, explained in
Section~\ref{sec:bethe}, where nodes are divided into generations.
Each node provides input to an ancestor node and receives input
from $c\!-\!1$ descendent nodes. This forms a tree structure, in
which the state of a node depends on those of its subsequent
generations.

In order to derive a set of recursive equations one should make
an assumption about the inherent symmetries of the problem.
Here we employ the replica symmetric ansatz. 
In previous treatment of related problems, 
the order parameters are represented as an integral 
over moments of the corresponding probability distribution, 
incorporating the permutation invariance 
of the replica indices~\cite{KSreview,wong87}. 
Generalizing to the case of $P_{\mathbf s}(\mathbf z)$, 
which is an order parameter depending on 
the continuous variables $\mathbf z$, 
the ansatz takes the form
\begin{equation}
        P_{\mathbf s}(\mathbf z)
        \!=\!\left\langle\prod_\alpha
        \left(\int d\nu~R(z_\alpha,\nu|{\mathbf T})
        \nu^{s_\alpha}\right)\right\rangle_\Lambda,
\label{moments}
\end{equation}
where ${\mathbf T}$ represents the tree terminated at the vertex node
with current potential $\nu$, providing input to the ancestor with
chemical potential $z$, and $\langle\dots\rangle_\Lambda$ represents
the average of the capacities of each node of the tree over the
distribution $\rho(\Lambda)$. Note that the replicas are coupled
through their common dependence on the quenched variables
$\Lambda$. This is in contrast to conventional derivations, such as
the SK model~\cite{Nishimori_book}, in which the dependence on the
disorder is integrated out, leading to more explicit inter-replica
dependencies.

The resultant recursion relation for the function
$R$ is independent of the replica indices, and hence remains valid
in the $n\to 0$ limit. It is given by
\begin{eqnarray}
        &&R(z,\nu \vert {\rm {\bf T}})=\frac{1}{\sD_R}
        \prod\limits_{k=1}^{c-1} \left[ {\int
        {d\nu _k R(\nu ,\nu _k \vert {\mathbf T}_k} )} \right]
        \nonumber\\
        &&\times\Theta \left( {\sum\limits_{k=1}^{c-1} {\nu _k } -c\nu +z
        +\Lambda_{V({\mathbf T})} } \right)
        \nonumber \\
        &&\times\exp \left[ {-\frac{\beta \varepsilon }{2}\nu ^2-\beta
        \sum_{k=1}^{c-1} {\phi \left( {\nu -\nu _k } \right)} } \right],
\label{eq:R_T}
\end{eqnarray}
where $\sD_R$ is a constant given in Eq.~(\ref{eq:R_TD}),
and ${\mathbf T}_k$ represents the tree terminated at
the $k$th descendent of the vertex.
$\Lambda _{V(T)}$ is the capacity of the vertex of the tree $\mathbf{T}$.
Eq.~(\ref{eq:R_T}) expresses $R(z,\nu|{\mathbf T})$ in terms of
$c\!-\!1$ functions $R(\nu,\nu_k|{\mathbf T}_k)$ ($k\!=\!1,..,c-1$),
integrated over $\nu_k$. Again, this is characteristic of the Bethe
lattice structure. Furthermore, except for the factor
$\exp(-\beta\epsilon\nu^2/2)$, 
a self-consistent solution of $R$ 
is that it is a function of 
$y\!\equiv\!\nu\!-\!z$, which is interpreted as the current drawn from
a node with current potential $\nu$ by its ancestor with current
potential $z$. 
Hence we will express the function $R$ as the product of
a {\em vertex partition function} $Z_V$ and a normalization factor $W$, 
which contains any residual dependence on $\nu$. 
Since $\epsilon$ is taken to approach zero in the analysis, 
it is expected that $W$ should approach a constant 
independent of $\nu$. 
Hence we let $R(z,\nu|{\mathbf T})\!=\!W(\nu)Z_V(y|{\mathbf T})$. 
As explained in Appendix~\ref{sec:app_replica}, 
in the limit $\epsilon\to 0$,
the dependence on $\nu$ and $y$ decouples; this enables one to derive
a recursion relation for the {\em vertex free energy}~\cite{footnote}
$F_V(y|{\mathbf T})\!\equiv\!-T\ln
Z_V(y|{\mathbf T})$ when a current $y$ is drawn from the vertex of a
tree ${\mathbf T}$ \cite{footnote1},
\begin{eqnarray}
        &&F_V(y|{\mathbf T})\!=\!-T\ln\Biggl\{
        \prod_{k\!=\!1}^{c-1}\left(\int dy_k\right)
        \Theta\!\Biggl(\sum_{k\!=\!1}^{c-1}y_k\!-\!y
        \nonumber\\
        &&+\Lambda_{V({\mathbf T})}\Biggr)
        \exp\left[-\beta\sum_{k\!=\!1}^{c-1}
        \left(F_V(y_k|{\mathbf
        T}_k)+\phi(y_k)\right)\right]\Biggr\}
        \nonumber\\
        &&+\Biggl\langle T\ln\Biggl\{ \prod_{k\!=\!1}^c\left(\int
        dy_k\right) \Theta\left(\sum_{k\!=\!1}^c y_k\!+\!\Lambda_V\right)
        \nonumber\\
        &&\times\exp\left[-\beta\sum_{k\!=\!1}^c \left(F_V(y_k|{\mathbf T}_k)
        \!+\!\phi(y_k)\right)\right]\Biggr\}
        \Biggr\rangle_\Lambda.
\label{recurt}
\end{eqnarray}
In the zero temperature limit, this recursion relation reduces to
\begin{eqnarray}
        &&F_V(y|{\mathbf T})
        \nonumber\\
        &&=\min_{\{y_k|\sum_{k\!=\!1}^{c-1}
        y_k-y+\Lambda_{V({\mathbf T})}\ge 0\}}
        \left[\sum_{k\!=\!1}^{c-1}
        \left(F_V(y_k|{\mathbf T}_k)+\phi(y_k)\right)\right]
        \nonumber\\
        &&-\left\langle\min_{\{y_k|\sum_{k\!=\!1}^c y_k+\Lambda_V\ge 0\}}
        \left[\sum_{k\!=\!1}^c
        \left(F_V(y_k|{\mathbf T}_k)+\phi(y_k)\right)\right]
        \right\rangle_\Lambda \ .
        \nonumber\\
\label{recur}
\end{eqnarray}
The solution of Eqs.~(\ref{recurt}) or (\ref{recur}) can be
obtained numerically in a recursive manner, since the vertex free
energy of a node depends on its own capacity and the disordered
configuration of its descendents.

Using the replica approach, and following the derivation
of Appendix~\ref{sec:app_replica}, the averaged free energy of the network
is given by
\begin{eqnarray}
        &&\langle F\rangle_{\Lambda}
        \!=\!-N\Biggl\langle T\ln\Biggl\{
        \prod_{k\!=\!1}^c\left(\int dy_k\right)
        \Theta\left(\sum_{k\!=\!1}^c y_k+\Lambda_V\right)
        \nonumber\\
        &&\times\exp\left[-\beta\sum_{k\!=\!1}^c
        \left(F_V(y_k|{\mathbf T}_k)+\phi(y_k)\right)\right]\Biggr\}
        \Biggr\rangle_\Lambda.
\label{free}
\end{eqnarray}

The current distribution and the average energy per link can
be derived, using the calculated vertex free energy, by
integrating the current $y'$ in a link from one vertex to another,
fed by the trees ${\mathbf T}_1$ and ${\mathbf T}_2$,
respectively; the obtained expressions are
$P(y)\!=\!\langle\delta(y-y')\rangle_{\star}$ and $\langle
\phi\rangle\!=\!\langle\phi(y')\rangle_{\star}$ where
\begin{equation}
        \langle\bullet\rangle_{\star}=
        \left\langle\frac
        {\int dy'\exp\left[-\beta F_L(y'|{\mathbf T}_1,{\mathbf T}_2)\right]
        (\bullet)}
        {\int dy'\exp\left[-\beta F_L(y'|{\mathbf T}_1,{\mathbf T}_2)\right]}
        \right\rangle_\Lambda,
\end{equation}
with
\begin{equation}
        F_L(y'|{\mathbf T}_1,{\mathbf T}_2)
        =F_V(y'|{\mathbf T}_1)+F_V(-y'|{\mathbf T}_2)+\phi(y').
\end{equation}

\section{Recursion Relation and Free Energy in the Bethe Approach}
\label{sec:bethe}

The results in Section \ref{sec:replica}
can be interpreted using the Bethe approach.
Since the connectivity $c$ is low, the probability of finding a
loop of finite length on the graph is low, and the Bethe
approximation well describes the local environment of a node. In
this approximation, a node is connected to $c$ branches in a tree
structure, and the correlations among the branches of the tree are
neglected. In each branch, nodes are arranged in generations. A
node is connected to an ancestor node of the previous generation,
and another $c-1$ descendent nodes of the next generation.

Consider a vertex $V({\mathbf T})$ of capacity
$\Lambda_{V({\mathbf T})}$, and a current $y$ is drawn from the
vertex. At a temperature $T\equiv\beta^{-1}$, one can write an
expression for the free energy $F(y|{\mathbf T})$ as a function of
the free energies $F(y_k|{\mathbf T}_k)$ of its descendants, that
branch out from this vertex
\begin{eqnarray}
        &&F(y|{\mathbf T})=-T\ln\Biggl\{
        \prod_{k=1}^{c-1}\left(\int dy_k\right)
        \Theta\!\left(\sum_{k=1}^{c-1} y_k\!-y+\!\Lambda_V\right)
        \nonumber\\
        &&\times\exp\left[-\beta\sum_{k=1}^{c-1}
        \left(F_V(y_k|{\mathbf T}_k)+\phi(y_k)\right)\right]\Biggr\},
\end{eqnarray}
where ${\mathbf T}_k$ represents the tree terminated at the
$k^{\rm th}$ descendent of the vertex. The free energy can be
considered as the sum of two parts
\[F(y|{\mathbf
T})\!=\!N_{\mathbf T}F_{\rm av}\!+\!F_V(y|{\mathbf T}) , \] where
$N_{\mathbf T}$ is the number of nodes in the tree ${\mathbf T}$,
$F_{\rm av}$ is the average free energy per node, and
$F_V(y|{\mathbf T})$ is the vertex free energy.

This allows one to decompose the free energy into
\begin{eqnarray}
        &&\left( {\sum\limits_{k=1}^{c-1} {N_{{\rm {\bf T}}_k
        } } +1} \right)F_{av} +F_V (y\vert {\rm {\bf
        T}})=\sum\limits_{k=1}^{c-1} {N_{{\rm {\bf T}}_k } F_{av} }
        \nonumber\\
        &&-T\ln \Biggl\{ \prod\limits_{k=1}^{c-1} {\left( {\int {dy_k } }
        \right)} \Theta \left( \sum\limits_{k=1}^{c-1} {y_k } -y+\Lambda
        _{V({\mathbf T})} \right)
        \nonumber\\
        &&\times\exp \left[ {-\beta
        \sum\limits_{k=1}^{c-1} {\left( {F_V (y_k \vert {\rm {\bf T}}_k
        )+\phi(y_k)} \right)} } \right] \Biggr\}.
\label{eq:freeBethe}
\end{eqnarray}
To determine the $F_{av}$, we consider the effects of adding a
vertex V which is fed by $c$ individual trees \textbf{T}$_{1}$,
{\ldots}, \textbf{T}$_{c}$. The total free energy is then
\begin{eqnarray}
\label{eq:freeBetheTot}
        &&\left( {\sum\limits_{k=1}^c {N_{{\rm {\bf
        T}}_k } } +1} \right)F_{av} = -\Biggl\langle T\ln \Biggl\{
        \prod\limits_{k=1}^c \left( \int {dy_k }  \right)
        \nonumber\\
        &&\Theta \left({\sum\limits_{k=1}^c {y_k } +\Lambda _V } \right)
        \exp \Biggl[ -\beta
        \sum\limits_{k=1}^c \biggl( N_{{\rm {\bf T}}_k } F_{av} +F_V (y_k
        \vert {\rm {\bf T}}_k )
        \nonumber\\
        &&+\phi(y_k) \biggr)  \Biggr] \Biggr\}
        \Biggr\rangle _\Lambda .
\end{eqnarray}

Rearranging the terms one obtains a recursion relation identical
to Eq.~(\ref{recurt}).
The average free energy per node is given
by the second term of Eq.~(\ref{recurt}), 
and has the same expression for the free energy as
in the replica approach (\ref{free}).

The recursion relation can also be recast into a form 
reminiscent of those commonly appearing in Bethe lattices 
of Ising spin variables, such as in Refs.~\cite{MZreview,KSreview,wong87}. 
This is achieved by considering the probability distribution 
of vertex free energies $P[F_V]$. 
Using Eq.~(\ref{recur}),
\begin{eqnarray}
	&&P[F_V]=\int d\Lambda_V\rho(\Lambda_V)\sum_{k=1}^{c-1}
	\int\cD F_{Vk}P[F_{Vk}]
	\nonumber\\
	&&\times\prod_y\delta\Biggl(-T\ln\Biggl\{
        \prod_{k=1}^{c-1}\left(\int dy_k\right)
        \Theta\!\left(\sum_{k=1}^{c-1} y_k\!-y+\!\Lambda_V\right)
        \nonumber\\
        &&\times\exp\left[-\beta\sum_{k=1}^{c-1}
        \left(F_{Vk}(y_k)+\phi(y_k)\right)\right]\Biggr\}
	\nonumber\\
	&&-\langle F\rangle_\Lambda-F_V(y)\Biggr).
\end{eqnarray}
Comparing with Bethe lattices of Ising variables, 
the vertex free energy $F_V$ plays the role of a cavity field. 
The difference here is that the distribution to be iterated 
is no longer a function of a single cavity variable. 
Rather, the distribution is defined 
in the space of cavity free energy {\em functions}, 
since we are dealing with continuous variables. 
This parallelism enables us to solve the recursion relation 
by {\em population dynamics}. 
At each step of this approach, a new ancestor node 
is generated at random, 
and its vertex free energy is updated.

It is interesting to point out that the iterative
equations~(\ref{recurt}) can be directly linked to those obtained from
a principled Bayesian approximation, where the logarithms of the
messages passed between nodes are proportional to the vertex free
energies. This is shown explicitly in Appendix~\ref{sec:app_Bayesnets}.

\section{The Message-passing Algorithm}
\label{sec:algorithm}

The local nature of the recursion relation~(\ref{recurt}) points to
the possibility that the network optimization can be solved by message
passing approaches, However, in contrast to other message passing
algorithms which pass conditional probability estimates of {\em
discrete values} to neighboring nodes, the messages in the present
context are more complex, since they are {\it functions}
$F_V(y|{\mathbf T})$ of the current $y$.

The derivation of the algorithm can be viewed as a minimization of
the cost function with respect to current changes under the
capacity constraint at the neighboring nodes. When the cost is
quadratic, the impact of current changes can be described through
the first and second derivatives with respect to the vertex free
energy.
As will be explained at the end of this section, 
this two-component message is sufficient to provide the exact solution, 
as long as the algorithm converges.

We follow this route and simplify the message to two parameters,
namely, the first and second derivatives of the vertex free
energies.
Let $(A_{ij},B_{ij})\equiv (\partial F_V(y_{ij}|{\mathbf
T}_j)/\partial y_{ij},
\partial^2 F_V(y_{ij}|{\mathbf T}_j)/\partial y_{ij}^2)$
be the message passed from node $j$ to $i$.
Based on the messages received from the descendents $k\ne i$,
the vertex free energy from $j$ to $i$ can be obtained
by minimizing the free energy
in the space of the current adjustments $\varepsilon_{jk}$
drawn from the descendents.
Using Eq.~(\ref{recur}), we minimize
\begin{equation}
        F_{ij}=\sum_{k\ne i}\cA_{jk}\left[A_{jk}\varepsilon_{jk}
        +\frac{1}{2}B_{jk}\varepsilon_{jk}^2
        +\phi'_{jk}\varepsilon_{jk}
        +\frac{1}{2}\phi''_{jk}\varepsilon_{jk}^2\right],
\label{fij}
\end{equation}
subject to the constraint
\begin{equation}
        \sum_{k\ne i}\cA_{jk}(y_{jk}+\varepsilon_{jk})-y_{ij}+\Lambda_j\ge 0,
\label{conj}
\end{equation}
where $\phi'_{jk}$ and $\phi''_{jk}$
represent the first and second derivatives of $\phi(y)$
at $y=y_{jk}$ respectively.
Introducing the Lagrange multiplier $\mu_{ij}$,
the optimal solution is given by
\begin{equation}
        F^*_{ij}=\frac{1}{2}\sum_{k\ne i}\cA_{jk}
        \left[\mu_{ij}^2-(A_{jk}+\phi'_{jk})^2\right]
        (B_{jk}+\phi''_{jk})^{-1}
\label{fopt}
\end{equation}
where
\begin{eqnarray}
        &&\mu_{ij}=\min\Biggl\{
        \Biggl[\sum_{k\ne i}\cA_{jk}[y_{jk}-(A_{jk}+\phi'_{jk})
        (B_{jk}+\phi''_{jk})^{-1}]
        \nonumber\\
        &&+\Lambda_j-y_{ij}\Biggr]
        \Biggl[\sum_{k\ne i}\cA_{jk}(B_{jk}+\phi''_{jk})^{-1}\Biggr]^{-1},
        0\Biggr\}.
\label{msgmu}
\end{eqnarray}
The first and second derivatives of $F^*_{ij}$ with respect to $y_{ij}$ 
lead to the forward message $(A_{ij},B_{ij})$ from node $j$ to $i$,
\begin{equation}
        A_{ij}\!\leftarrow\!-\mu_{ij},
        \quad
        B_{ij}\!\leftarrow\!\frac{\Theta(-\mu_{ij}+\epsilon)}
        {\sum_{k\ne i}\cA_{jk}(B_{jk}+\phi''_{jk})^{-1}}.
\label{msgab}
\end{equation}
We note in passing that
when the descendent currents $y_{jk}$ change continuously,
both the vertex free energy~(\ref{fopt})
and the chemical potential~(\ref{msgmu}) change continuously
for functions $\phi(y)$ with continuous first derivatives.
Hence for the quadratic load balancing task,
defined by $\phi(y)=y^2/2$,
a self-consistent solution of the recursion relation Eq.~(\ref{recur})
consists of vertex free energies
which are piecewise quadratic with continuous slopes.
This makes the 2-parameter message a very precise approximation.

In principle, if the forward messages
consist of the full vertex free energy functions,
then they are already sufficient for the optimization task.
However, since the messages are simplified
to be the first and second derivatives of the vertex free energies,
each node needs to estimate the optimal currents by
approximating the vertex free energy function as a quadratic function.
Hence, the remaining step of the algorithm
is the determination of the drawn current $y_{ij}$
at which the derivatives comprising the messages should be computed.
This determination of the {\it working point}
is achieved by passing additional information-provision messages
among the nodes,
a step not present in conventional message-passing algorithms.
The following two methods are proposed for this purpose.

In the first method, when messages are sent
from node $j$ to ancestor node $i$,
backward messages $y_{jk}$ computed from the same optimization steps
are sent from node $j$ to the descendent nodes $k\ne i$,
informing them of the particular arguments
to be used for calculating subsequent messages.
From Eqs.~(\ref{fij}) and (\ref{conj}),
this backward message is given by
\begin{equation}
        y_{jk}\!\leftarrow\!y_{jk}-\frac{
        A_{jk}+\phi'_{jk}+\mu_{ij}}{
        B_{jk}+\phi''_{jk}}.
\label{msgback}
\end{equation}

In the second method, node $j$ first
receives the messages $(A_{ji},B_{ji})$ and current $y_{ji}$
from the ancestor node $i$,
and update the current from $y_{ij}$ to $y_{ij}+\varepsilon_{ij}$
by minimizing the total cost
\begin{eqnarray}
        &&E_{ij}=A_{ij}\varepsilon_{ij}+\frac{1}{2}B_{ij}\varepsilon_{ij}^2
        +A_{ji}(-y_{ij}-\varepsilon_{ij}-y_{ji})
        \nonumber\\
        &&+\frac{1}{2}B_{ji}(-y_{ij}-\varepsilon_{ij}-y_{ji})^2
        +\phi'_{ij}\varepsilon_{ij}+\frac{1}{2}\phi''_{ij}\varepsilon_{ij}^2.
\label{eij}
\end{eqnarray}
In Eq.~(\ref{eij}), the first two terms
represent the message from $i$ to $j$,
the next two terms from $j$ to $i$,
and the last two terms the transportation cost in link $(ij)$.
The optimal solution is
\begin{equation}
        y_{jk}\!\leftarrow\!\frac{
        B_{ij}y_{ij}-A_{ij}-B_{ji}y_{ji}+A_{ji}
        -\phi'_{ij}+\phi''_{ij}y_{ij}}{
        B_{ij}+B_{ji}+\phi''_{ij}}.
\label{infoanc}
\end{equation}
Both methods work well for the quadratic cost function.

Implicit in the information-provision messages
is the independent update of the currents $y_{ij}$ and $y_{ji}$
in the opposite directions of the same link.
This allows us to use the criterion $y_{ij} = -y_{ji}$
as a check for the convergence of the algorithm.
We have used this in our simulation
program by requiring the root mean square average of $y_{ij}+ y_{ji}$
to be less than a threshold.
Another usage of the information-provision messages
is in monitoring the optimal cost function during simulations.  This
saves the extra step of calculating the current associated with a link
in the conventional Bethe approach.

An alternative distributed algorithm can be obtained
by iterating the chemical potentials of the nodes.
The K\"uhn-Tucker condition requires
that the terms $\mu_i(\sum_j\cA_{ij}y_{ij}+\Lambda_i)$
in Eq.~(\ref{lagr}) vanish.
Eliminating $y_{ij}$ in terms of the chemical potentials,
$\mu_i$ can be expressed in terms of $\mu_j$ of its neighbors, namely,
\begin{equation}
        \mu_i=\min(g_i^{-1}(0),0);
        \quad
        g_i(x)=\sum_j\cA_{ij}[\phi']^{-1}(\mu_j-x)+\Lambda_i.
\label{mufunc}
\end{equation}
For the quadratic load balancing task, $\phi(y)=y^2/2$ and
\begin{equation}
        \mu_i=\min\left[\frac{1}{c}\left(
        \sum_j\cA_{ij}\mu_j+\Lambda_i\right),0\right].
\label{itermu}
\end{equation}
This provides a local iteration method for the optimization problem.
We may interpret this algorithm as a price iteration scheme,
by noting that the Lagrangian in Eq.~(\ref{lagr}) can be written as
\begin{equation}
        L=\sum_{(ij)}\cA_{ij}L_{ij}+{\rm constant},
\end{equation}
where
\begin{equation}
        L_{ij}=\phi(y_{ij})+(\mu_i-\mu_j)y_{ij}.
\end{equation}
Therefore, the problem can be decomposed into independent optimization
problems, each for a current on a link.  $\mu_i$ is the storage price
at node $i$, and each subproblem involves balancing the transportation
cost on the link, and the storage cost at node $i$ less that at node
$j$, yielding the optimal solution.  This provides a pricing scheme
for the individual links to optimize, which simultaneously optimizes
the global performance~\cite{kelly}.

It can be easily verified that the message-passing algorithm, 
in the two-parameter approximation, 
yield solutions identical to the price iteration algorithm, 
which is exact, 
as long as the connectivity is sparse 
and the algorithms converge. 
Indeed, simulations provided in section~\ref{sec:numerical} 
show that the two algorithms yield excellent agreement with each other.

One can proceed with the verification 
by noting from Eq.~(\ref{msgback}) 
that $\mu_{ij}=-\phi'_{jk}-A_{jk}$ for all $k\ne i$ 
at the steady state. 
Since $\mu_{ij}$ is independent of the node $i$ 
receiving the message, 
one can write $\mu_{ij}$ as $\mu_j$. 
Similarly, using Eq.~(\ref{msgab}), $A_{jk}=-\mu_{jk}=-\mu_k$. 
We then have $\phi'_{jk}=\mu_k-\mu_j$, 
whose inverse relation is Eq.~(\ref{yij}). 
The resource constraint Eq.~(\ref{constr}) 
then leads to Eq.~(\ref{mufunc}). 

The result that the first order message 
converges to the exact result of the chemical potential 
at the steady state 
justifies the simplification of the message to two parameters. 
It illustrates that higher order messages 
are not required for the precision of the optimal solution, 
as long as the algorithm converges. 
This is natural for the quadratic cost, 
for which it can be verified that the vertex free energies 
are piecewise quadratic functions of the currents 
with continuous slopes. 
In addition, exact solutions can be found for other cost functions, 
as described in Section~\ref{sec:generalcost}. 
Though the second order messages do not play a role 
in the final solution, 
they are useful in tuning the intermediate steps for faster convergence. 
The situation is reminiscent of the use 
of both gradient and curvature information 
in many gradient-based optimization techniques.

\section{The High Connectivity Limit}
\label{sec:asymptotic}

Both the recursive~(\ref{recur}) and message-passing
equations~(\ref{msgmu})-(\ref{msgab}) can be solved numerically as will
be shown in the next section. However, scaling laws of the quantities
of interest can also be derived analytically in the limit of high
connectivity.

We restrict the analysis to the case of quadratic cost function
$\phi(y)=y^{2}/2$. 
In the limit of large $c$, Eq.~(\ref{msgab}) converges to the result
\begin{equation}
\label{eq:bK}
    B_{ij} =\frac{\Theta (-\mu _{ij} )}{c} \ ,
\end{equation}
and the currents scale as $c^{-1}$.
Therefore, the task of satisfying the capacity constraints
is shared by a fraction of $O(1)$ of the descendents.
As a result, the collective effects of the descendents on a node
can be expressed in terms of the statistical properties of the descendents.
Using this scaling property of the currents,
Eq.~(\ref{msgab}) reduces to
\begin{equation}
\label{eq:aK}
        A_{ij} =\max \left( {\frac{1}{c}\left[ {\sum\limits_{k\ne i}
        {\cA_{jk} A_{jk}} -\Lambda _j } \right],0} \right).
\end{equation}
By virtue of the law of large numbers, it is sufficient to consider the
mean $m_A$ and variance $\sigma_A^2$ of the messages
$A_{ij}$. Respectively, they scale as $c^{-1}$ and $c^{-2}$ with $\sum
_{k\ne i}\cA_{jk}A_{jk}$ being self-averaging. Hence, we can write
\begin{equation}
\label{eq:aK1}
        A_{ij} =\frac{1}{c}(cm_A -\Lambda _j )\Theta (cm_A -\Lambda _j ).
\end{equation}

Averaging over $\Lambda$, drawn from a Gaussian of mean
$\langle \Lambda \rangle $ and variance 1 used in our numerical
studies, one obtains a self-consistent expression for the parameter
$\xi\equiv cm_A-\langle \Lambda \rangle$
\begin{equation}
\label{eq:xiK}
        \left\langle \Lambda \right\rangle
        =\int_{-\infty }^\xi {Dz(\xi -z)} -\xi
        =\frac{e^{-\frac{\xi ^2}{2}}}{\sqrt {2\pi } }-\xi H(\xi ).
\end{equation}

\subsection{Current distribution}

To obtain the current distribution,
one considers the vertex free energies of both ends of a link.
For a current $y_{ij}$ flowing from $j$ to $i$,
the total energy is given by
\begin{equation}
\label{eq:EK}
        E=A_{ij} y_{ij} +\frac{\Theta (-\mu _{ij} )}{2c}y_{ij}^2
        -A_{ji} y_{ij}
        +\frac{\Theta (-\mu _{ji} )}{2c}y_{ij}^2 +\frac{1}{2}y_{ij}^2,
\end{equation}
where we have approximated the working points of the messages
to be $y_{ij}=0$.
This is justified since the magnitudes of the messages are $O(c^{-1})$,
and $y_{ij}\sim c^{-1}$.
Minimizing the energy with respect to the current $y_{ij}$, one finds
\begin{eqnarray}
\label{eq:yK}
        &&y_{ij} =\frac{1}{c}\bigl[(cm_A-\Lambda_i)\Theta(cm_A-\Lambda_i)
        \nonumber\\
        &&-(cm_A-\Lambda_j)\Theta(cm_A-\Lambda_j)\bigr].
\end{eqnarray}
Hence, the current distribution is given by
\begin{eqnarray}
        &&P(y)=\int d\Lambda_1\rho(\Lambda_1)
        \int d\Lambda _2\rho(\Lambda_2)
        \delta\Biggl\{\frac{1}{c}\biggl|(cm_A-\Lambda_1)
        \nonumber\\
        &&\times\Theta(cm_A -\Lambda_1)
        -(cm_A-\Lambda_2)\Theta(cm_A-\Lambda_2)\biggr|
        -y\Biggr\}.
    \nonumber\\
\label{eq:yKscale}
\end{eqnarray}
For the Gaussian distribution of capacities, one obtains
\begin{eqnarray}
        &&P(y)=2\frac{\exp\left({-\frac{c^2y^2}{4}}
        \right)}{\sqrt {4\pi /c^2} }H\left( {\frac{cy-2\xi }{\sqrt 2 }}
        \right)
        \nonumber\\
        &&+2H(\xi )\frac{\exp \left( {-\frac{(cy-\xi )^2}{2}}
        \right)}{\sqrt {2\pi /c^2} }+H(\xi )^2\delta (y) \ .
\end{eqnarray}
This shows that the distribution $P(cy)/c$, obtained by
rescaling the argument by $c^{-1}$, is independent of $c$, and depends
solely on the average capacity $\langle \Lambda \rangle $ through
$\xi $. In particular, the fraction of {\it idle} links is given by
\begin{equation}
\label{eq:y_idle}
        P(y=0)=H(\xi)^2.
\end{equation}
The physical picture of this scaling behavior
is that the total current required by a node
to satisfy its capacity constraint
is shared by the links.

\subsection{Average energy}

Using Eq.~(\ref{eq:yK}), the average energy per link can be written as
\begin{eqnarray}
\label{eq:EKscale}
        &&\left\langle\phi\right\rangle
        =\frac{1}{c^2}\biggl\{ \left\langle
        {(cm_A-\Lambda)^2\Theta(cm_A-\Lambda)}\right\rangle
        \nonumber\\
        &&-\left\langle
        {(cm_A-\Lambda)\Theta(cm_A -\Lambda)}\right\rangle^2\biggr\}.
\end{eqnarray}
For the Gaussian capacity distribution, it becomes
\begin{equation}
\label{eq:EKgauss}
        \langle\phi\rangle
        =\frac{1}{c^2}\left[I_2(\xi)-I_1(\xi)^2\right],
\end{equation}
where
\begin{equation}
\label{eq:I1}
        I_1 (\xi )=\int_{-\infty }^\xi {Dz(\xi -z)} =\frac{e^{-\frac{\xi
        ^2}{2}}}{\sqrt {2\pi } }+\xi H(-\xi ),
\end{equation}
\begin{equation}
\label{eq:I2}
        I_2 (\xi )=\int_{-\infty }^\xi {Dz(\xi -z)^2}
        =\xi \frac{e^{-\frac{\xi^2}{2}}}{\sqrt {2\pi } }
        +(\xi ^2+1)H(-\xi ).
\end{equation}

We are also interested in how the energy is distributed 
in the network. 
Consider the average energy per link $\langle\phi|\Lambda\rangle$ 
among those links connected to nodes of capacity $\Lambda$. 
Using Eq.~(\ref{eq:yK}), 
\begin{eqnarray}
        &&\left\langle\phi|\Lambda\right\rangle
        =\frac{1}{c^2}\int d\Lambda_2\rho(\Lambda_2)\biggl[
        (cm_A-\Lambda)\Theta(cm_A-\Lambda)
        \nonumber\\
        &&(cm_A-\Lambda_2)\Theta(cm_A -\Lambda_2)\biggr]^2.
\end{eqnarray}
For the Gaussian capacity distribution, this becomes
\begin{equation}
\label{eq:enode}
        \langle\phi|\Lambda\rangle
        =\frac{1}{2c^2}\left[I_2(\xi)-\left(I_1(\xi)^2-\Lambda^2\right)
	\Theta\left(I_1(\xi)-\Lambda\right)\right].
\end{equation}

\subsection{Chemical potential distribution}

To obtain the distribution of the chemical potentials, one follows a
similar treatment and considers a central node 0 fed by $c$
descendents.  Introducing a Lagrange multiplier to enforce the
capacity constraint, one replaces the energy minimization problem
by the Lagrangian
\begin{eqnarray}
\label{eq:Lagrange_mu}
        &&L=\sum\limits_{j=1}^c {\left[ {A_{0j} y_{0j} +\frac{\Theta(-\mu_{0j}
        )}{2c}y_{0j}^2 +\frac{1}{2}y_{0j}^2 } \right]}
        \nonumber\\
        &&+\mu \left(
        {\sum\limits_{j=1}^c {y_{0j} } +\Lambda _0 } \right).
\end{eqnarray}
The currents are given by $y_{0j} =-A_{0j} -\mu$, and the chemical
potential by
\[
        \mu =\min \left( {\frac{1}{c}\left[ {\Lambda _0 -\sum\limits_{j=1}^c
        {A_{0j} } } \right],0} \right).
\]
In the large $c$ limit, the approximated expression for $\mu$ becomes
\begin{equation}
\label{eq:muK}
        \mu =\frac{1}{c}\left({\Lambda_0-cm_A}\right)\Theta\left({cm_A
        -\Lambda _0}\right),
\end{equation}
and the chemical potential distribution is similarly derived
\begin{eqnarray}
\label{eq:PmuK}
        P(\mu)=&&\int_{-\infty}^{cm_A}d\Lambda~\rho(\Lambda)\delta\left[
        \frac{1}{c}\left({\Lambda-cm_A}\right)-\mu\right]
        \nonumber\\
        &&+\int_{cm_A}^\infty d\Lambda~\rho(\Lambda)\delta(\mu).
\end{eqnarray}
For the Gaussian capacity distribution, it reduces to
\begin{equation}
\label{eq:PmuKgauss}
        P(\mu )=\frac{\exp \left( {-\frac{(c\mu +\xi )^2}{2}} \right)}{\sqrt
        {2\pi /c^2} }\Theta (-\mu )+H(\xi )\delta (\mu ).
\end{equation}
This shows that the distribution $P(c\mu)/c$, obtained by rescaling
the argument by $c^{-1}$, is independent of $c$, and depends solely on
the average capacity $\langle \Lambda \rangle $ through $\xi$. In
particular, the fraction of unsaturated nodes is given by
\begin{equation}
        P(\mu=0)=H(\xi).
\end{equation}

\subsection{Resource Distribution}

We define the resource at a node $i$ by
\begin{equation}
        r_i\equiv\Lambda_i+\sum_j\cA_{ij}y_{ij}.
\end{equation}
The currents are obtained by minimizing
\begin{equation}
        E_i=\sum_j\cA_{ij}\left[A_{ij}y_{ij}+\frac{\Theta(-\mu_j)}{2c}
        y_{ij}^2+\frac{1}{2}y_{ij}^2\right]
\end{equation}
subject to the constraints $\sum_j\cA_{ij}y_{ij}+\Lambda_i\ge 0$.
Introducing the Lagrange multiplier $\mu_i$ for the constraint,
we obtain
\begin{equation}
        r_i=\max[\Lambda_i-cm_A,0].
\end{equation}
Hence, the resource distribution is given by
\begin{equation}
        P(r)=\int^\infty_{cm_A}\!\!d\Lambda~\rho(\Lambda)
        \delta(\Lambda-cm_A-r)
        +\int^{cm_A}_{-\infty}\!\!d\Lambda~\rho(\Lambda)\delta(r).
\end{equation}
For the Gaussian capacity distribution, it reduces to
\begin{equation}
        P(r)=\frac{\exp\left[-\frac{1}{2}(r-\xi)^2\right]}{\sqrt{2\pi}}
        \Theta(r)+H(-\xi)\delta(r).
\end{equation}
This shows that the resource distribution becomes independent of $c$
in the large $c$ limit,
confirming the picture that the current in a link
scales as $c^{-1}$,
summing up to a total current of $c^0$
satisfying the resource requirement of the nodes.

\subsection{Dynamics}

To analyze the dynamics in the limit of large $c$,
one considers random sequential updates
using the algorithm presented in section~\ref{sec:algorithm}.
Time is divided into steps of size $\Delta t = 1/cN$.
At each time step, a directed link from node $j$ to $i$
is randomly chosen,
such that each directed link is chosen exactly once
in each integer interval of time,
and the messages of the links are updated.

The current $y_{jk}$, for a link feeding node $j$, is updated in
the backward messages corresponding to the forward ones from $j$
to $i \ne k$. (This implies that $y_{jk}$ is updated $K$ times in
a time step. As will be shown, the algorithm uses information
updated in the previous step to compute the optimal current. Since
the previous step lies in the previous interval, this approach is
not the most efficient for monitoring the evolving average
energy.)

Denote the average of message $A_{jk}(t)$ over all links at time
$t$ as $m_A(t)$, and $\hat{m}_A (t)$ the expected value of
$A_{jk}(t)$ when it is updated at time $t$. Then, for a time $t$
in the interval between $t_{0}$ and $t_{0}+ 1$  this leads to the
dynamical equation
\begin{equation}
\label{eq:dma}
        \frac{dm_A(t)}{dt}=\hat {m}_A(t)-m_A(t_0 )
        \mbox{ for } t_{0}\le t <t_{0}+1 \ .
\end{equation}

Suppose the link $ij$ is updated at time $t$,
according to Eq.~(\ref{msgab}).
Then the average over link $ij$ becomes
\begin{equation}
\label{eq:maK}
        c\hat{m}_A(t)=\int_{-\infty }^{cm_A(t)}
        d\Lambda~\rho(\Lambda)~(cm_A(t)-\Lambda) \ .
\end{equation}
For the Gaussian capacity distribution, this becomes
\begin{equation}
\label{eq:maKGauss}
        c\hat {m}_A(t)=\int_{-\infty}^{\xi(t)}
        Dz~(\xi(t)-z)=I_1(\xi(t)) \ ,
\end{equation}
where $\xi(t)= cm_A(t)-\langle\Lambda\rangle$.
It is convenient to convert Eq. (\ref{eq:dma})
to a dynamical equation for $\xi(t)$
\[
        \frac{d\xi (t)}{dt}=I_1 (\xi (t))-\xi (t_0 )-\left\langle \Lambda
        \right\rangle  \mbox{ for } t_{0}\le t< t_{0} + 1 \ ,
\]
with the initial condition $\xi(0) = -\langle \Lambda \rangle $.

The dynamics of the average energy depends on
whether one adopts the backward or forward information-provision method,
described by Eqs.~(\ref{msgback}) and (\ref{infoanc}) respectively.
We first consider the case of backward information-provision.
Suppose the link from $j$ to $i$
is updated at time $t_{ij}$
in the interval between $t_0$ and $t_0+1$.
Using Eq.~(\ref{msgback}),
\begin{equation}
\label{eq:yback}
        y_{jk}(t_{ij})=-A_{jk}(t_{jk}^-)-\mu _{ij}(t_{ij}),
\end{equation}
where $A_{jk}(t_{jk}^-)$ is given by Eq.~(\ref{eq:aK1})
and $t_{jk}^{-}$ is the instant
that the link from $k$ to $j$
is previously chosen for update.
With probability $t_0+1-t_{ij}$,
$t_{jk}^-$ lies in the previous time interval
between $t_0-1$ and $t_0$.
Otherwise, $t_{jk}^-$ lies between $t_0$ and $t_{ij}$.

To calculate the average energy $\langle\phi(t_0+1)\rangle$,
one can express $\langle y_{jk}^2\rangle$
in terms of the moments
$\langle A_{jk}^2(t_{jk}^-)\rangle$,
$\langle\mu_{ij}^2(t_{ij})\rangle$ and
$\langle A_{jk}(t_{jk}^-)\mu_{ij}(t_{ij})\rangle$.
Hence, on averaging over all descendents $k$,
denoted as $\langle\ \rangle_k$, 
the second moment of $A_{jk}(t_{jk}^-)$ is given
via Eq.~(\ref{eq:aK1}) by
\begin{eqnarray}
        &&\left\langle A_{jk}(t_{jk}^-)^2\right\rangle_k
        \nonumber\\
        &&=\frac{1}{c^2}\left[\int_{t_0-1}^{t_0}dt_{jk}^-(t_0+1-t_{ij})
        +\int_{t_0}^{t_{ij}}dt_{jk}^-\right]
        \\
        &&\times\int d\Lambda~\rho(\Lambda)
        \left[cm_A(t_{jk}^-)-\Lambda\right]^2
        \Theta\left[cm_A(t_{jk}^-)-\Lambda\right].
        \nonumber
\label{eq:Amoments}
\end{eqnarray}
For the Gaussian capacity distribution, this becomes
\begin{equation}
        \left\langle A_{jk}(t_{jk}^-)^2\right\rangle_k
        =\frac{1}{c^2}\overline I_2(t_{ij}),
\end{equation}
where
\begin{equation}
        \overline I_l(t)
        \equiv\left[\int_{t_0-1}^{t_0}dt'(t_0+1-t)
        +\int_{t_0}^t dt'\right]I_l(\xi(t')).
\end{equation}
Averaging over $(ij)$ at time $t_0+1$,
\begin{equation}
        \left\langle A_{jk}(t_{jk}^-)^2\right\rangle_{ijk}
        =\frac{1}{c^2}\int_{t_0}^{t_0+1}dt~\overline I_2(t),
\end{equation}
which can be simplified to
\begin{eqnarray}
        &&\left\langle A_{jk}(t_{jk}^-)^2\right\rangle_{ijk}
        =\frac{1}{c^2}\Biggl[\frac{1}{2}\int_{t_0-1}^{t_0}dt~I_2(\xi(t))
        \nonumber\\
        &&+\int_{t_0}^{t_0+1}dt~(t_0+1-t)I_2(\xi(t))\Biggr].
\end{eqnarray}

A similar calculation follows for the second moment of
$\mu_{ij}(t_{ij})$ at time $t_0+1$, leading to
\begin{equation}
\label{eq:mumoments}
        \langle\mu_{ij}(t_{ij})^2\rangle_{ij}
        =\frac{1}{c^2}\int_{t_0}^{t_0+1}dt~I_2(\xi(t)),
\end{equation}
for the Gaussian capacity distribution.

Similarly, the expression for the crossed moment
in the case of Gaussian capacity distribution is
\begin{equation}
        \langle A_{jk}(t_{jk}^-)\mu_{ij}(t_{ij})\rangle_{ijk}
        =-\frac{1}{c^2}\int_{t_0}^{t_0+1}dt~\overline I_1(t)I_1(\xi(t)),
\end{equation}
which can be simplified to
\begin{eqnarray}
        &&\langle A_{jk}(t_{jk}^-)\mu_{ij}(t_{ij})\rangle_{ijk}
        =-\frac{1}{c^2}\Biggl\{
        \left[\int_{t_0-1}^{t_0}dt~I_1(\xi(t))\right]
        \nonumber\\
        &&\times\left[\int_{t_0}^{t_0+1}dt~(t_0+1-t)I_1(\xi(t))\right]
        \nonumber\\
        &&+\frac{1}{2}\left[\int_{t_0}^{t_0+1}dt~I_1(\xi(t))\right]^2
        \Biggr\}.
\end{eqnarray}
Hence the average energy per link
in the case of Gaussian capacity distribution is
\begin{eqnarray}
        &&\langle\phi(t_0+1)\rangle
        =\frac{1}{2c^2}\Biggl\{
        \frac{1}{2}\int_{t_0-1}^{t_0}dt~I_2(\xi(t))
        \nonumber\\
        &&+\int_{t_0}^{t_0+1}dt~(t_0+1-t)I_2(\xi(t))
        -2\left[\int_{t_0-1}^{t_0}dt~I_1(\xi(t))\right]
        \nonumber\\
        &&\times\left[\int_{t_0}^{t_0+1}\!\!dt~(t_0+1-t)I_1(\xi(t))\right]
        \!-\!\left[\int_{t_0}^{t_0+1}\!\!dt~I_1(\xi(t))\right]^2
        \nonumber\\
        &&+\int_{t_0}^{t_0+1}dt~I_2(\xi(t))\Biggr\}.
\label{eq:muGauss}
\end{eqnarray}

Next, we consider the case of forward information-provision
described by Eq.~(\ref{infoanc}),
whose large $c$ limit is given by
\begin{equation}
        y_{ij}(t_{ij})=-A_{ij}(t_{ij}^-)+A_{ji}(t_{ji}^-),
\label{forhik}
\end{equation}
where the link from $j$ to $i$ is updated at time $t_{ij}$
in the time interval between $t_0$ and $t_0+1$,
$t_{ij}^-$ and $t_{ji}^-$ are respectively the instants
that the links from $j$ to $i$ and from $i$ to $j$
are previously chosen for update.
Derivation analogous to
the backward information-provision method yields
\begin{eqnarray}
        &&\langle\phi(t_0+1)\rangle
        =\frac{1}{2c^2}\Biggl\{
        \int_{t_0-1}^{t_0}dt~I_2(\xi(t))
        -2\biggl[\int_{t_0-1}^{t_0}dt
        \nonumber\\
        &&\times I_1(\xi(t))\biggr]
        \left[\int_{t_0}^{t_0+1}\!\!dt~\overline I_1(t)\right]
        +\int_{t_0}^{t_0+1}\!\!dt~\overline I_2(t)\Biggr\},
\end{eqnarray}
which can be simplified to
\begin{eqnarray}
        &&\langle\phi(t_0+1)\rangle
        =\frac{1}{2c^2}\Biggl\{
        \frac{3}{2}\int_{t_0-1}^{t_0}dt~I_2(\xi(t))
        \\
        &&+\int_{t_0}^{t_0+1}\!\!dt~(t_0+1-t)I_2(\xi(t))
        -2\left[\int_{t_0-1}^{t_0}\!\!dt~I_1(\xi(t))\right]
        \nonumber\\
        &&\times\left[\int_{t_0}^{t_0+1}\!\!dt~(t_0+1-t)I_1(t)\right]
        -\left[\int_{t_0-1}^{t_0}\!\!dt~I_2(\xi(t))\right]^2\Biggr\}.
        \nonumber
\end{eqnarray}
Eq.~(\ref{forhik}) shows that
the forward information-provision method
uses only outdated information to calculate the current.
Consequently, the convergence of the average energy
is slower than that of the backward information-provision method
by about half a step.

\section{Numerical Results}
\label{sec:numerical}

To examine the accuracy of the theoretical results and the
efficacy of the message passing algorithm of
section~\ref{sec:algorithm} we conducted a series of numerical
experiments. First, we solved numerically the recursive equation
(\ref{recur}) using population dynamics 
for various connectivity values and obtained various
quantities of interest from it, including the energy, current and
chemical potential distribution. Second, we carried out simulations
using the algorithms of Eqs.~(\ref{msgab}) and (\ref{msgmu}) and
compared the results to those obtained from the numerical
simulations. We then compared the scaling properties of the
results with respect to the connectivity with the theoretical
scaling obtained in section~\ref{sec:asymptotic}. All experiments
in this section have been carried out using the quadratic
cost $\phi(y)=y^{2} /2$.

To solve numerically the recursive equation (\ref{recur}) we have
discretized the vertex free energy functions $F_V(y|{\mathbf T})$
into a vector, whose $i^{\rm th}$ component is the value of the
function corresponding to the current $y_i$. To speed up the
optimization search at each node, we first find the {\it vertex
saturation current} drawn from a node such that:
(a) the current drawn by each descendent node
separately optimizes its own vertex free energy
plus the transportation cost to the node being considered.
For descendent nodes $k$,
this current $y_k^*$ is given by
\begin{equation}
        y_k^*={\rm arg min}_{y_k}\left[F_V(y_k|{\mathbf T}_k)
        +\phi(y_k)\right].
\end{equation}
(b) The resource of the node considered is just used up.
For node $j$, its vertex saturation current $y^{s}_{j}$ is given by
\begin{equation}
        y^{s}_{j}=\sum_{k\ne j}\cA_{jk}y_k^*+\Lambda_j.
\end{equation}
For current below this saturation point,
the vertex free energy remains constant.
That is, $F_V(y_j|{\mathbf T}_j)=F_V(y^{s}_{j}|{\mathbf T}_j)$
for $y_j\le y^{s}_{j}$.
Hence, this provides a convenient starting point
for searching the optimal solutions.
The drawn current can then be increased in small discrete steps,
and the optimal solution can be found, for example,
using an exhaustive search,
by varying the descendent currents in small discrete steps.
This approach is particularly convenient for $c=3$,
where the search is confined to a single parameter.
For larger values of $c$, other search techniques,
such as conjugate gradient search, can be used.

These recursive equations provide a discretized representation
of the vertex free energy $F_V(y|{\mathbf T})$,
from which various properties of the system can be calculated.

{\em Average Energy -} To compute the average energy, we randomly
draw $c-1$ nodes, compute the optimal current flowing between them,
and repeat the sampling to obtain the average.

The results of iteration for a Gaussian distribution $\rho(\Lambda)$ 
with variance 1 and average $\langle\Lambda\rangle$ 
were described in \cite{us_prerc}. 
There we found that the convergence rate slows down 
when $\langle\Lambda\rangle$ decreases towards 0. 
A cusp in the relaxation rate dependence on $\langle\Lambda\rangle$ 
exists at $\langle\Lambda\rangle\approx 0.45$, 
where the fraction of saturated nodes is about 0.48, 
close to the percolation threshold of 0.5 for $c=3$. 
Hence the cusp is probably related to the appearance 
of a percolating cluster of negative resources, 
which draws currents from increasingly extensive regions of nodes 
with excess resources to satisfy the demand.

{\em Dependence on the connectivity -} 
We have presented in \cite{us_prerc} 
evidence that the currents scale as $c^{-1}$, 
so that after rescaling, 
the average energy $c^2\langle\phi\rangle$ (see Fig. 1 inset), 
the current distribution $P(cy)/c$, 
and the resource distribution $P(r)$ 
become principally dependent 
on the average capacity $\langle\Lambda\rangle$, 
and only weakly dependent on the connectivity $c$.
The scaling property extends to the dynamics of the optimization process. 
All results of increasing connectivity approach 
those of the high connectivity limit 
derived in Section~\ref{sec:asymptotic}.

Here we further study how the high connectivity limit is approached 
from increasing finite values of $c$. 
We define the empirical scaling factor $s$ by
\begin{equation}
	s=\sqrt{\frac{\lim_{c\to\infty}c^2\langle\phi\rangle}
	{\langle\phi\rangle}}.
\end{equation}
$s$ is expected to converge to $c$ in the high conenctivity limit. 
As shown in Fig. 1, 
the empirical scaling factor corresponding 
to different values of $\langle\Lambda\rangle$ 
collapse on a linear curve with slope 1. 
The best fit is $s\approx 1.02c-0.43$. 
It is remarkable that the network behavior 
already converges to scaling at low values of $c$.

\begin{figure}[t]
\begin{center}
\begin{picture}(200,360)
\put(0,180){\epsfxsize=70mm  \epsfbox{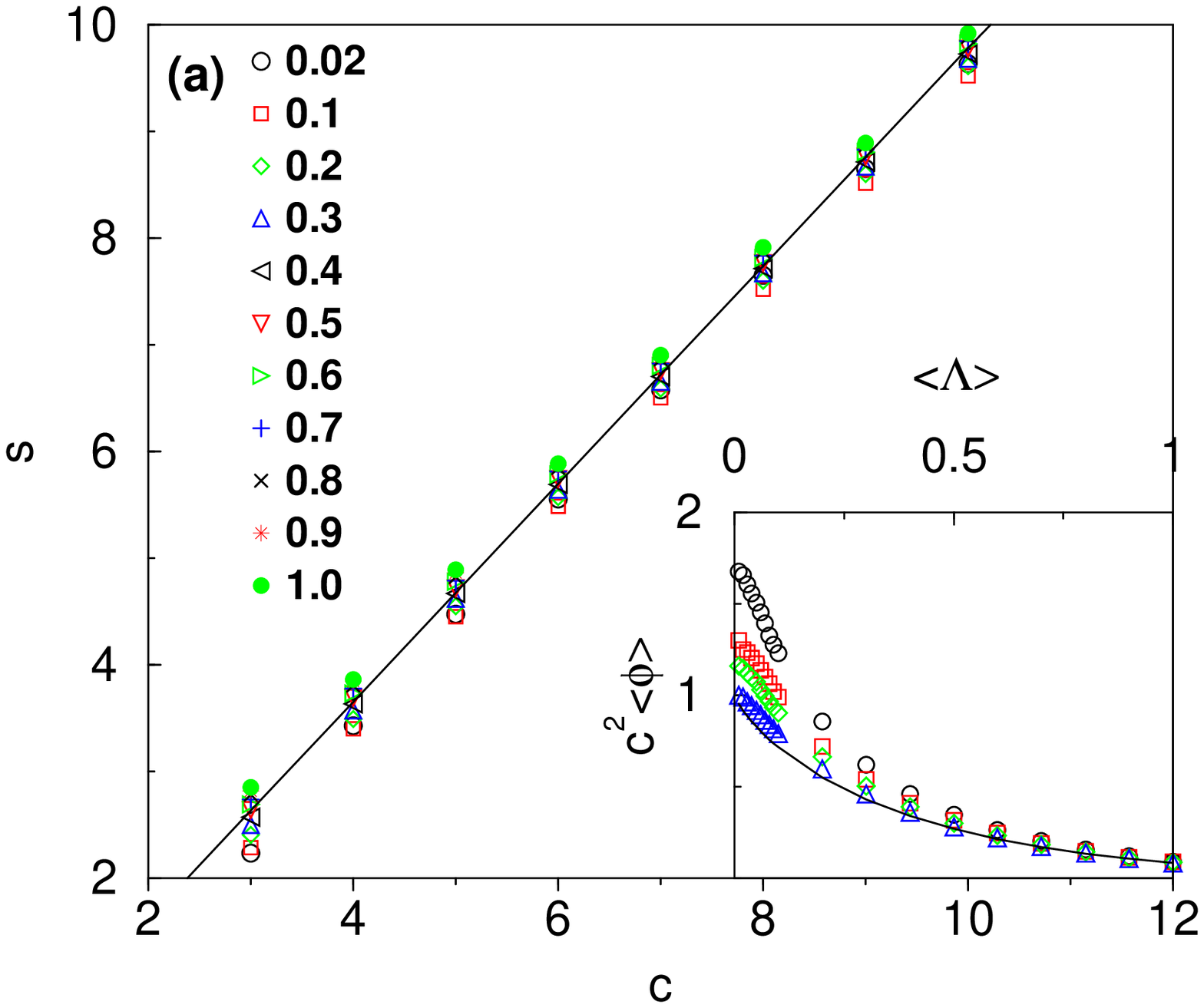}}
\put(0,0){\epsfxsize=70mm  \epsfbox{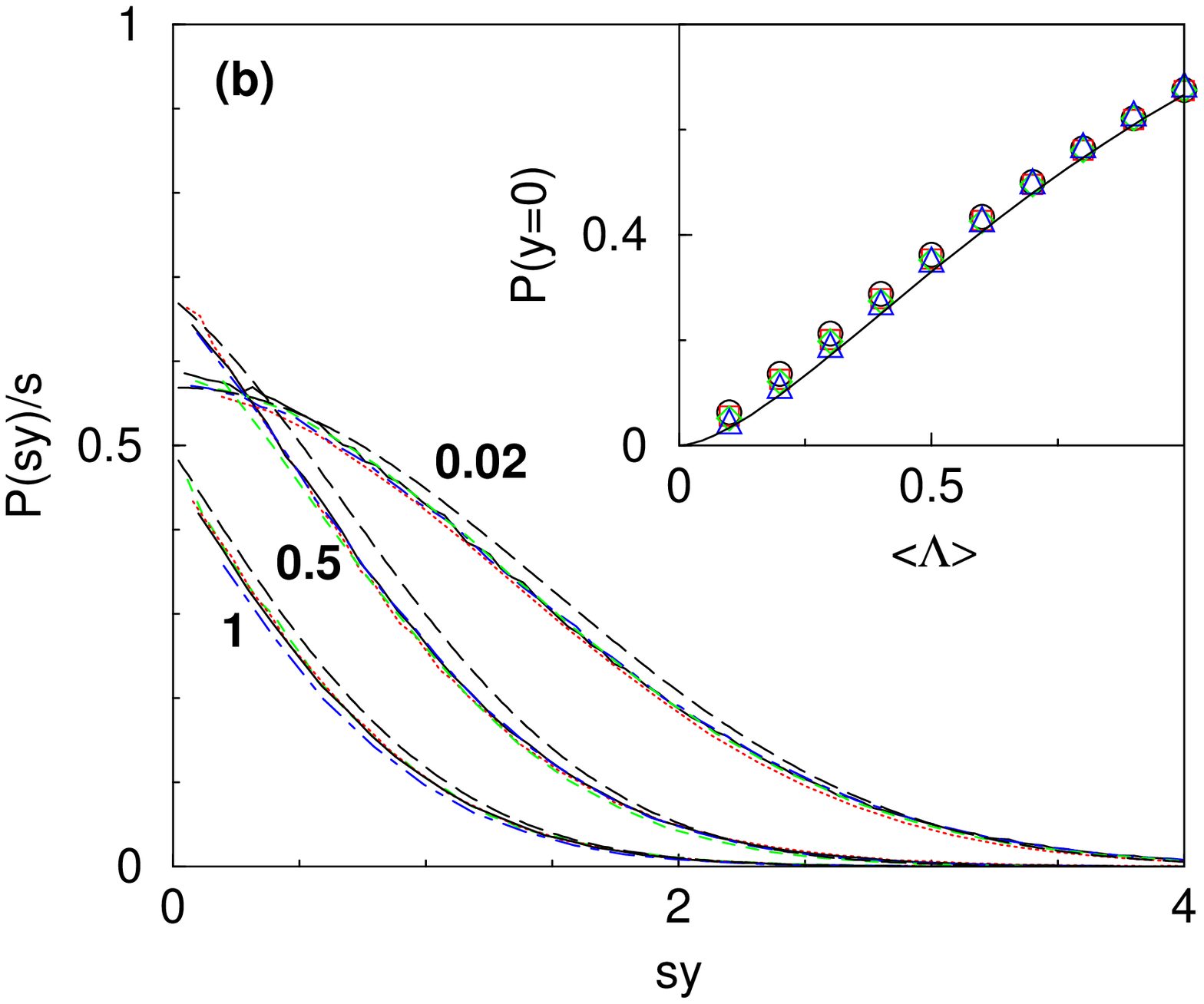}}
\end{picture}
\caption{
(color online) 
Results for $N\!=\!1000$ and $\phi(y)=y^2/2$. 
(a) The dependence of the empirical scaling factor $s$ 
on the connectivity $c$ for different values of $\langle\Lambda\rangle$. 
Line: best fit. 
Inset: $c^2\langle\phi\rangle$ as a function of $\langle\Lambda\rangle$. 
Symbols: $c$=3 ($\bigcirc$), 4 ($\square$), 5 ($\lozenge$), 
10 ($\bigtriangleup$), high $c$ (line).
(b) The continuous component of the current distribution $P(sy)/s$ 
for $\langle \Lambda\rangle=0.02, 0.5, 1$. 
Lines: $c=3$ (solid), 4 (dotted), 5 (dashed), 10 (dot-dashed), 
high $c$ (long dashed). 
Inset: $P(y\!=\!0)$ as a function of $\langle\Lambda\rangle$, 
symbols: same as (a) inset.
\vspace*{-0.5cm}}
\label{fig:1}
\end{center}
\end{figure}

We make use of this empirical scaling factor 
to rescale the current distribution. 
The current distribution consists of a delta function component 
at $y=0$ (Fig. 1(b) inset~\cite{footnote2}) and a continuous component, 
whose breadth decreases with $\langle\Lambda\rangle$. 
Excluding the delta function component, 
the continuous distribution after rescaling is shown in Fig. 1(b). 
The approach to the high connectivity limit is even faster 
when compared with that by setting the scaling factor to be $c-1$ 
\cite{us_prerc}.

{\em Energy dependence on node capacity -}
We divide the nodes into ten groups according to their node capacities. 
Nodes in group 1 have their capacities among the top 10\%, 
those in group 2 the next highest 10\%, and so on. 
For each group, we then calculate the average energy per link 
for those links connected to the nodes of that group. 
The general trend can be seen in Figs. 2(a-b). 
Group 1 consists of the richest nodes. 
Since they are the resource providers to the rest of the network, 
their connected links have high average energy. 
On the other hand, group 10 consists of the poorest nodes. 
Since they are the resource consumers of the network, 
their connected links also have high average energy. 
Compared with group 1, 
their average energy is even higher 
due to the enforcement of the resource constraints, Eq.~(\ref{constr}). 
By comparison, the groups in between consist of relay nodes 
which typically receive resources from the richer ones 
and provide resources to the poorer ones.
The energies of their connected links have intermediate averages. 
Figs. 2(a-b) show that these different roles 
played by nodes of different capacities 
can lead to a significant difference 
in the average energies of their connected links.

\begin{figure}[t]
\begin{center}
\begin{picture}(200,360)
\put(0,180){\epsfxsize=70mm  \epsfbox{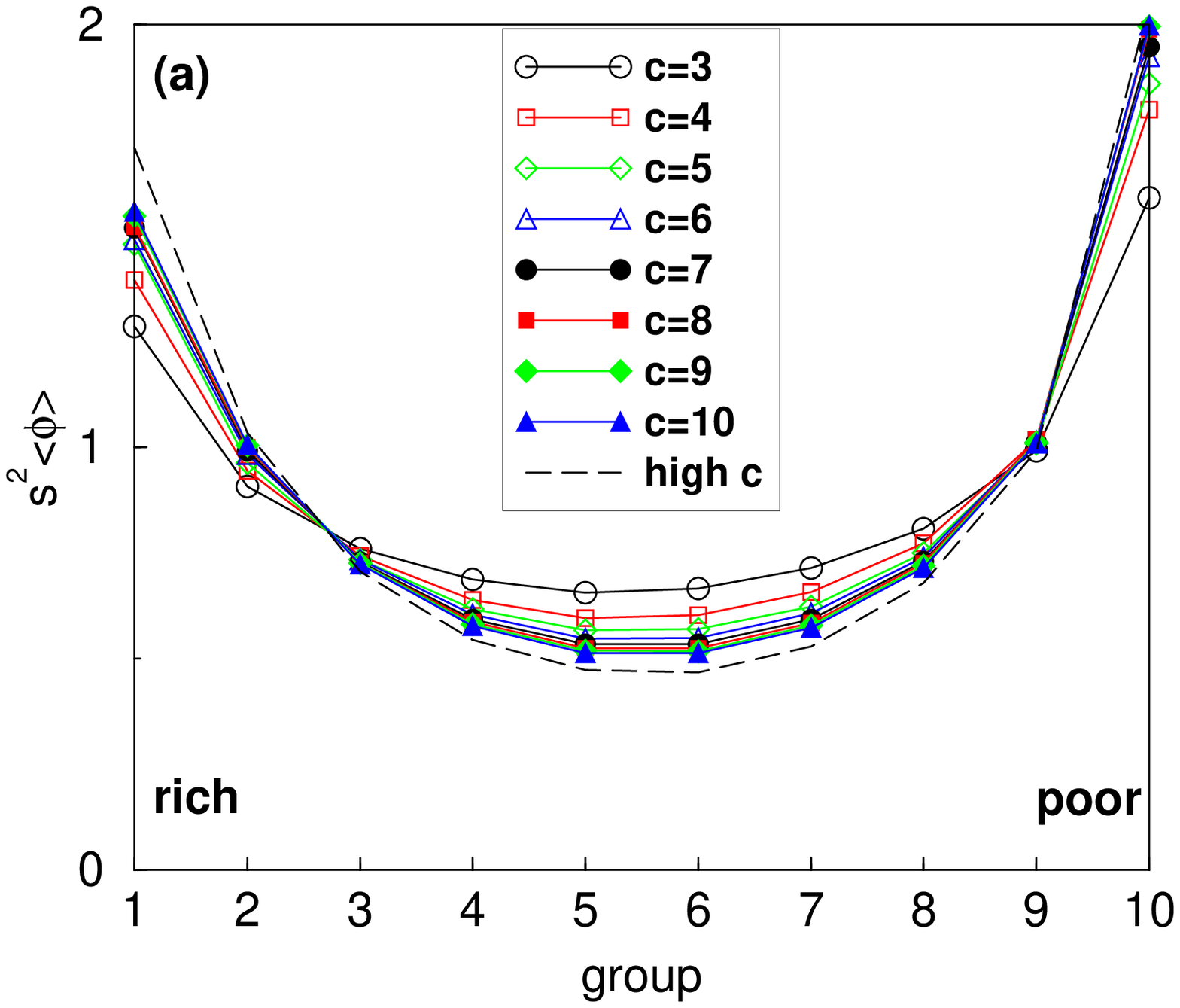}}
\put(0,0){\epsfxsize=70mm  \epsfbox{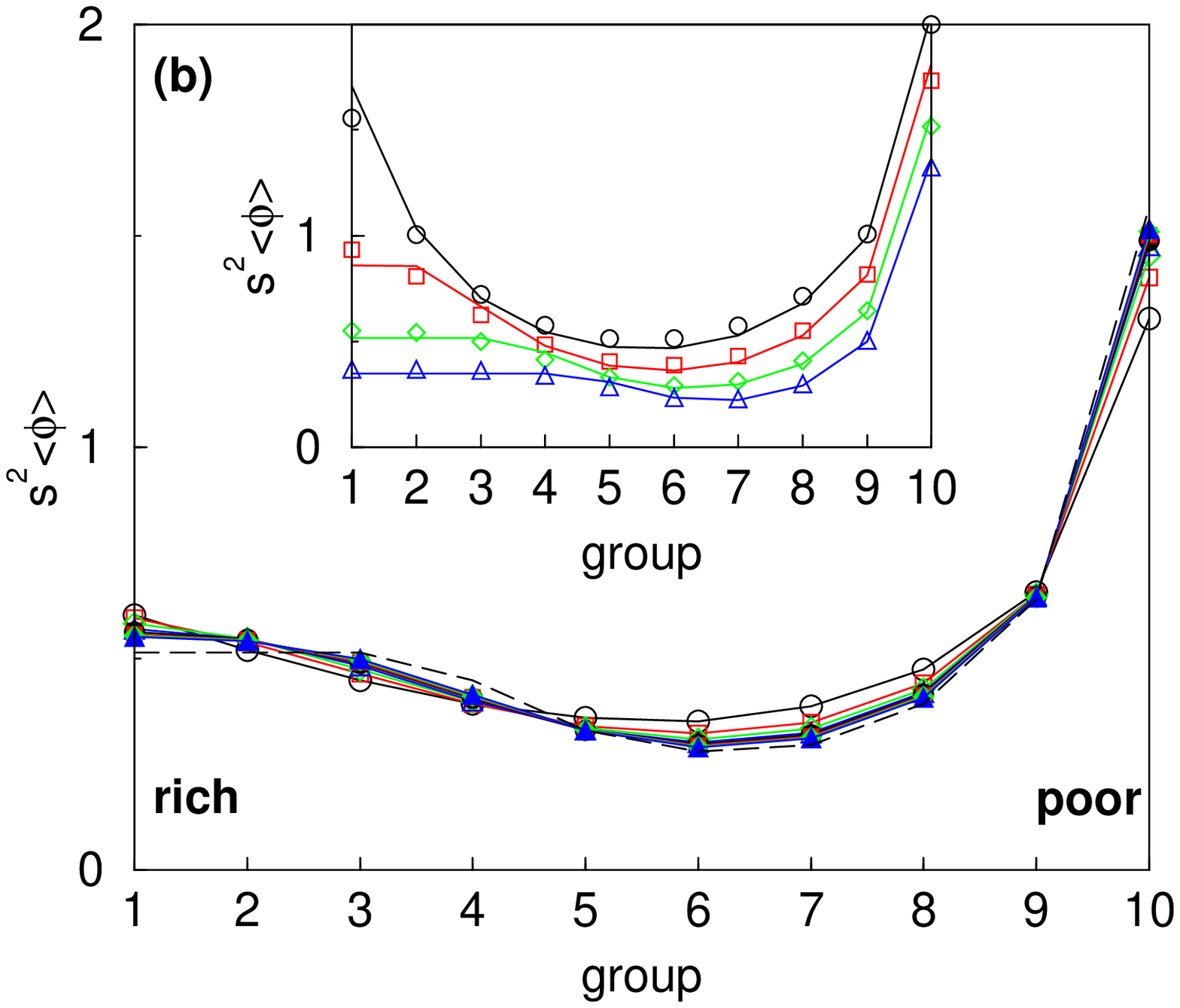}}
\end{picture}
\caption{
(color online) Results for $N\!=\!1000$ and $\phi(y)=y^2/2$.
(a) The rescaled energy per link 
connected to the ten groups of nodes 
with decreasing ranges of capacities 
at $\langle\Lambda\rangle=0.02$ 
for different connectivities indicated in the legend. 
(b) Same as (a) but at $\langle\Lambda\rangle=0.2$. 
Inset: Comparison of the curves for 
$\langle\Lambda\rangle$=0.02, 0.1, 0.2, 0.3 (top to bottom). 
Lines: high $c$.
Symbols: $c=10$ and 
$\langle\Lambda\rangle$=0.02 ($\bigcirc$), 0.1 ($\square$), 0.2 ($\lozenge$), 
0.3 ($\bigtriangleup$).
\vspace*{-0.5cm}} \label{fig:2}
\end{center}
\end{figure}

Furthermore, when one compares networks of different connectivities, 
one finds that the rescaled energy curves become 
only weakly dependent on the connectivity. 
The convergence to the high connectivity limit is rather fast.

Fig.~3(b) inset shows the rescaled energy curves 
in the high connectivity limit 
for different $\langle\Lambda\rangle$. 
When $\langle\Lambda\rangle$ increases, 
a plateau starts to develop among the groups of richer nodes, 
indicating that the rich nodes become unsaturated 
in their resource provision, 
so that the energy of their connected links 
become independent of their excess resources. 
They have $\Lambda>I_1(\xi)$ according to Eq.~(\ref{eq:enode}).
Simulation results for $c=10$ presented in the same figure 
provide support to this behavior. 
In fact, the development of this plateau 
is already visible in the simulation results 
of finite connectivities in Fig. 3(b), 
whose trend shows that the homogenization of energy 
among the links connected to the rich nodes is increasingly effective 
when the connectivity increases.

\begin{figure}[t]
\begin{center}
\begin{picture}(200,180)
\put(0,0){\epsfxsize=70mm  \epsfbox{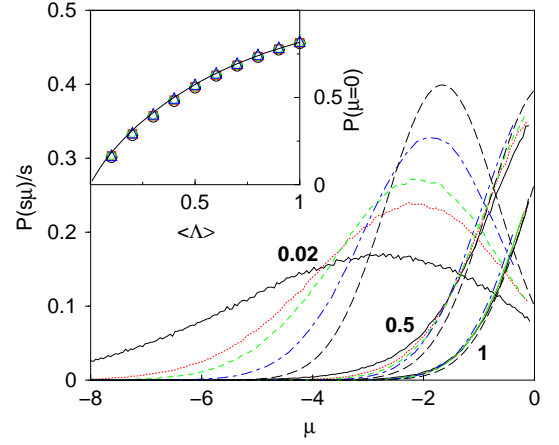}}
\end{picture}
\caption{
(color online) Results for system size $N\!=\!1000$ and $\phi(y)=y^2/2$.
The chemical potential distribution $P(s\mu)/s$
for $\langle\Lambda\rangle = 0.02, 0.5, 1$. 
Lines: $c=3$ (solid), 4 (dotted), 5 (dashed), 10 (dot-dashed), 
high $c$ (long dashed). 
Inset: $P(\mu=0)$ as a function of $\langle\Lambda\rangle$. 
Symbols: $c=3$ ($\bigcirc$), 4 ($\square$), 5 ($\lozenge$), 
10 ($\bigtriangleup$), high $c$ (long dashed).
\vspace*{-0.5cm}}
\label{fig:3}
\end{center}
\end{figure}

{\em Chemical potential distribution -}
Both the message-passing and price iteration algorithms
allow us to study the distribution $P(\mu)$
of the chemical potentials $\mu$.
$P(\mu)$ consists of a delta function at $\mu=0$ (Fig. 3 inset) 
and a continuous component. 
The width of the continuous component increases
when $\langle\Lambda\rangle$ decreases.
Note the concurrence of low average resource
and highly negative values of $\mu$,
consistent with the economic interpretation of $\mu$
as the storage cost of a node.
The scaling property of the distribution
is illustrated in Fig.~\ref{fig:3}.
For $\langle\Lambda\rangle$ = 0.5 and 1,
the distributions collapse well
even for relatively low values of $c$.
For $\langle\Lambda\rangle$ = 0.02,
a considerable dependence on $c$ remains after rescaling.
However, the approach to the high $c$ limit is visible.

\section{General Cost Functions}
\label{sec:generalcost}

The cost used so far was the, rather simple, quadratic cost. In
this section we examine the applicability of the message-passing
algorithm for more general costs. Two representative costs will be
studied:

(a) The {\it anharmonic} cost function is used to model the
effects of costs rising faster than quadratic
\begin{equation}
\label{eq:anharmoniccost}
        \phi (y)=\frac{y^2}{2}+\frac{u\left| y\right|^3}{3}.
\end{equation}

(b) The {\it frictional} cost function is used to model the effects of
frictional forces in resource allocation, which add an extra cost
per unit current in a link irrespective of direction. 
Hence it is also useful in networks 
with multiple classes of traffic sharing the same links. 
The cost function takes the form
\begin{equation}
\label{eq:frictionalcost}
        \phi (y)=\frac{y^2}{2}+v\left| y\right|.
\end{equation}
Note that these cost functions represent two distinct cases.
The former has well defined first and second derivatives for all
of its arguments.
In the latter case, the frictional cost function
does not have a second derivative at $y=0$.
There is a kink in the cost function
at the point of zero current,
thus increasing the preference for idle links,
or equivalently the reluctance to switch on a link.
As will be shown, the convergence of the message-passing algorithm
is much more difficult,
and modifications are necessary.

\subsection{Anharmonic cost}

\subsubsection{Price iteration}

Introducing Lagrange multipliers for the capacity constraints, the function
to be minimized is
\begin{equation}
\label{eq:anharmonicCost}
        L=\sum\limits_{(ij)} {\left(
        {\frac{y_{ij}^2 }{2}+\frac{u\left| {y_{ij} } \right|^3}{3}}
        \right)} +\sum\limits_i {\mu _i \left( {\sum\limits_{j\in \cL_i }
        {y_{ij} } +\Lambda_i } \right)} ,
\end{equation}
where $\cL_i$ is the set of neighboring nodes of $i$.
Optimizing with respect to $y_{ij}\equiv-y_{ji}$,
one obtains the relation
\begin{equation}
\label{eq:anharmonicSol}
        y_{ij} =\frac{1}{u}\left[ {\sqrt
        {\frac{1}{4}+u\left| {\mu _j -\mu _i } \right|} -\frac{1}{2}}
        \right]{\rm sgn}(\mu _j -\mu _i ).
\end{equation}
Using the capacity constraints, the chemical potential $\mu_{i}$
is given by $\mu_{i}= \min(g_{i}^{-1}(0), 0)$, where $g_{i}^{-1}$
is the inverse of the function
\begin{equation}
\label{eq:anharmonicg}
        g_i (x)=\sum\limits_{j\in \cL_i }
        {\frac{1}{u}\left[ {\sqrt {\frac{1}{4}+u\left| {\mu _j -x}
        \right|} -\frac{1}{2}} \right]{\rm sgn}(\mu _j -x)} +\Lambda_i .
\end{equation}
This provides a price iteration scheme.
We solve this equation using the bisection method,
noting that the function is monotonic non-increasing. This requires
one to know the solution bounds. Let $\mu_{\max}$ and $\mu_{\min}$
be the maximum and minimum of the chemical potentials among the
neighbors of node $i$. Examining the cases of $\Lambda _{i} > 0$
and $\Lambda _{i} < 0$ separately, one finds the range for the
solution of $x$
\begin{eqnarray}
\label{eq:anharmonic_muSol}
        &&\mu _{\min } +\min \left[
        {\frac{\Lambda_i }{c}\left( {1+\frac{u\left| {\Lambda_i }
        \right|}{c}} \right),0} \right]\le x\le
    \nonumber\\
    &&\min \left\{ {\mu _{\max }
        +\max \left[ {\frac{\Lambda _i }{c}\left( {1+\frac{u\left|
        {\Lambda_i } \right|}{c}} \right),0} \right],0} \right\}.
\end{eqnarray}

\subsubsection{Message-passing}
\label{sec:anharmonic_MP}

Since the cost function has well defined first and second
derivatives for all of its arguments, the message-passing
algorithm formulated in section~\ref{sec:algorithm} is directly
applicable:
\begin{equation}
\label{eq:anharmonic_a}
    A_{ij} \leftarrow -\mu_{ij} ,
\end{equation}
\begin{equation}
\label{eq:anharmonic_b}
    B_{ij} \leftarrow \frac{\Theta (-\mu _{ij}
    )}{\sum\limits_{k\in \cL_j \backslash \{i\}} (B_{jk}+\phi _{jk}''
    )^{-1} },
\end{equation}
where
\begin{eqnarray}
    &&\mu _{ij} =\min \Biggl\{
    \Biggl[\sum_{k\in \cL_j \backslash\{i\}}[y_{jk}-(A_{jk}+\phi'_{jk})
    (B_{jk}+\phi _{jk}'')^{-1}]
    \nonumber\\
    &&+\Lambda _j -y_{ij}
    \Biggr]\Biggl[
    \sum_{k\in \cL_j\backslash\{i\}}(B_{jk}+\phi _{jk}'')^{-1}\Biggr]^{-1}
    ,0 \Biggr\},\
\label{eq:anharmonic_mu}
\end{eqnarray}
and the backward message is given by
\begin{equation}
    y_{jk} \leftarrow y_{jk} -\frac{A_{jk}+\phi_{jk}'+\mu _{ij}}
    {B_{jk}+\phi _{jk}''}.
\label{eq::anharmonic_back}
\end{equation}
For the anharmonic cost function, \[\phi_{jk}'=y_{jk}+ u y_{jk}^2
~{\rm sgn}~y_{jk}, \mbox{ and }\phi_{jk}''= 1 + 2 u\vert y_{jk}\vert~. \]

\subsubsection{Simulation results}

To study the behavior of the various algorithms in the case the
anharmonic cost function, in comparison to the quadratic cost, we
carried out simulation under similar conditions to those of
section~\ref{sec:numerical}.

\begin{figure}
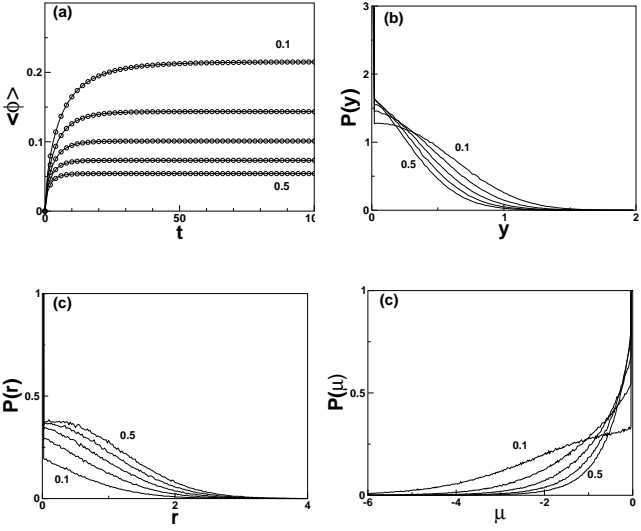

\centering
\begin{picture}(240,100)
\includegraphics[width=0.495\linewidth]{./fig4aM.eps}~
\includegraphics[width=0.46\linewidth]{./fig4bM.eps}
\end{picture}
\vgap
\begin{picture}(240,100)
\includegraphics[width=0.475\linewidth]{./fig4cM.eps}~
\includegraphics[width=0.475\linewidth]{./fig4dM.eps}
\end{picture}
\caption{Results for $N\!=\!1000$, $c\!=\!3$, the anharmonic cost
function with $u=1$, and 1000 samples. (a) $\langle\phi\rangle$
obtained by the price iteration algorithm, utilizing
Eqs.~(\ref{eq:anharmonicg}) and (\ref{eq:anharmonic_muSol}), as a
function of $t$ for $\langle \Lambda\rangle\!=\! 0.1, 0.2, 0.3,
0.4, 0.5$ (top to bottom). (b) The corresponding current
distribution $P(y)$. (c) The corresponding resource distribution
$P(r)$. (d) The corresponding chemical potential distribution
$P(\mu)$.} \label{fig:anharmonic}
\end{figure}

Figure~\ref{fig:anharmonic}(a) shows the average energy per link
as a function of iteration steps of the price iteration algorithm
for the anharmonic cost function.
Figures~\ref{fig:anharmonic}(b)-(d)
show the distributions, $P(y)$, $P(r)$ and $P(\mu)$
of the currents, resources and chemical potentials respectively,
at the corresponding values of $\langle\Lambda\rangle$.
The results obtained are very similar to those of the quadratic
cost function and show the same qualitative behavior,
as can be seen by comparing Figs.~\ref{fig:anharmonic}(a)-(d)
with Figs.~\ref{fig:1}(a) inset, \ref{fig:2}(a), \ref{fig:2}(b)
and \ref{fig:3}(a) respectively.

\begin{figure}[t]
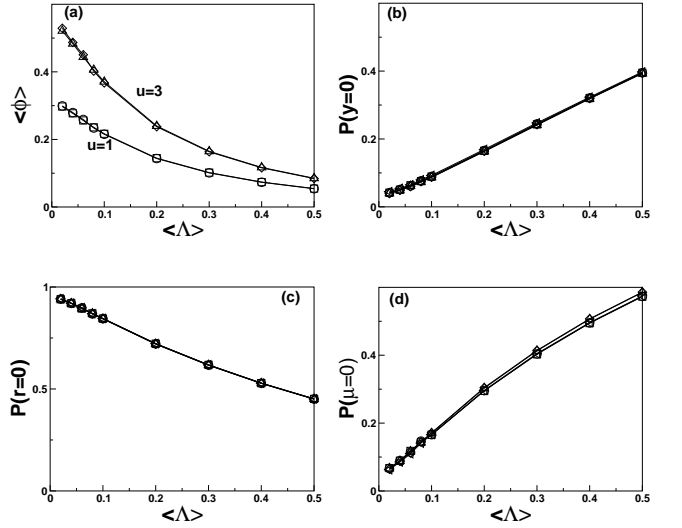

\centering
\begin{picture}(240,100)
\includegraphics[width=0.48\linewidth]{./fig5aM.eps}~
\includegraphics[width=0.48\linewidth]{./fig5bM.eps}
\end{picture}
\vgap
\begin{picture}(240,100)
\includegraphics[width=0.48\linewidth]{./fig5cM.eps}~
\includegraphics[width=0.48\linewidth]{./fig5dM.eps}
\end{picture}
\caption{Results for $N\!=\!1000$, $c\!=\!3$,
the anharmonic cost function, and 1000 samples.
(a) Average energy per link $\langle\phi\rangle$.
(b) The fraction of idle links $P(y=0)$.
(c) The fraction of saturated nodes $P(r=0)$.
(d) The fraction of unsaturated nodes $P(\mu=0)$.
Symbols in (a)-(d): results obtained
by the price iteration algorithm
for the quadratic cost function ($u=0$) ($\lhd$),
and the anharmonic cost function
with $u=1$ ($\bigcirc$) and $u=3$ ($\lozenge$);
results obtained by the message-passing algorithm
are also shown for $u=1$ ($\square$) and $u=3$ ($\bigtriangleup$).}
\label{fig:anharmonic_comp}
\end{figure}

In Figs.~\ref{fig:anharmonic_comp}(a)-(d)
we compare the behavior of the price iteration
and message-passing algorithms
by plotting the average energy per link,
the fraction of idle links $P(y=0)$,
the fraction of saturated nodes $P(r=0)$,
and the fraction of unsaturated nodes $P(\mu=0)$ respectively,
as a function of $\langle\Lambda\rangle$
for both algorithms
at the anharmonic strengths $u=1$ and 3.
Both methods converge to the same value throughout the range
examined and for the two $u$ values examined.

To provide a more quantitative comparison
of the cost functions,
we also plotted in
Figs.~\ref{fig:anharmonic_comp}(b)-(d) $P(y=0)$, $P(r=0)$ and $P(\mu=0)$
respectively, for the quadratic ($u= 0$) cost function.
Simulations have been carried
out under the same conditions ($N= 1000$, $c=3$ and 1000
samples). It is remarkable that there is little
difference between the quadratic and anharmonic cases.
The different cost functions
merely change the continuous components of these distributions,
leaving the delta function components effectively unchanged.

\subsection{Frictional cost}

\subsubsection{Price iteration}

Introducing Lagrange multipliers for the capacity constraints, the function
to be minimized is
\begin{equation}
\label{eq:frictionalCost}
        L=\sum\limits_{(ij)} {\left(
        {\frac{y_{ij}^2 }{2}+v\left| {y_{ij} } \right|} \right)}
        +\sum\limits_i {\mu _i \left( {\sum\limits_{j\in \cL_i } {y_{ij} }
        +\Lambda _i } \right)} .
\end{equation}
Optimizing with respect to $y_{ij} \equiv -y_{ji}$, one obtains
\begin{equation}
\label{eq:friction_y}
        y_{ij} =\left[ {\mu _j -\mu _i
        -v~\mbox{sgn}~(\mu _j -\mu _i )} \right]\Theta \left[ {\left| {\mu
        _j -\mu _i } \right|-v} \right].
\end{equation}
Using the capacity constraints, the chemical potential is given by
$\mu_{i} = \min~(g_{i}^{-1}(0), 0)$, where $g_{i}^{-1}$ is the
inverse of the function
\begin{equation}
\label{eq:g_friction}
        g_i (x)\!=\!\!\sum\limits_{j\in \cL_i } {\left[ {\mu
        _j -x-v ~\mbox{sgn}~(\mu _j -x)} \right]
    \!\Theta\! \left[ {\left| {\mu
        _j -x} \right|-v} \right]} +\Lambda _i  \ .
\end{equation}
Since $g_{i}(x)$ is monotonic non-increasing, and piecewise
linear, we have a fast way to solve the equation by finding the
function at its $2c$ turning points, located at $x= \mu_{j} \pm
 v$. If $g_{i} (\mu_{\min}- v)<0$, the solution is given by $(\mu_{\min}\!
- v)-g_{i}(\mu_{\min}\!-v)/c$;  if $g_{i} (\mu_{\min}\! - v) \ge
0$, then among the turning points with $g_{i}(x) \ge 0$, one finds
the one with the minimum value of $g_{i}(x)$, and the solution is
given by $x-g_{i}(x)/g_{i}'(x^+)$.

\subsubsection{Message-passing}

Message-passing algorithms for the currents have not been
successful in this case, presumably due to the divergence of the
second derivative at $y= 0$. This in turn requires some form of
regularization that causes the effects of friction to be exhibited
in the first, but not the second, derivative in finite systems.
This inconsistency prevents the algorithm from converging.

We present here an approach based on the chemical potential
representation. To formulate an appropriate version of
message-passing for this case, we return to the minimization of
the energy of section~\ref{sec:algorithm}, namely,
\begin{eqnarray}
\label{eq:energy_friction}
        &&E_j^{\backslash i} =\sum\limits_{k\in
        \cL_j \backslash \{i\}} \biggl[ A_{jk} \varepsilon _{jk}
        +\frac{1}{2}B_{jk} \varepsilon _{jk}^2 +\frac{1}{2}(y_{jk}
        +\varepsilon _{jk} )^2
        \nonumber\\
        &&+v\left| {y_{jk} +\varepsilon _{jk} }
        \right| \biggr] ,
\end{eqnarray}
subject to the constraints
\begin{equation}
\label{eq:constraints_friction}
        \sum\limits_{k\in \cL_j \backslash
        \{i\}} {(y_{jk} +\varepsilon _{jk} )} -y_{ij} +\Lambda _j \ge 0 \ ,
\end{equation}
introduced by employing the Lagrange multipliers $\mu _{ij}$. The
optimal solution is given by $\mu_{ij}= \min [g_{ij}^{-1}(0), 0]$,
where $g_{ij}^{-1}$ is the inverse of the function
\begin{eqnarray}
        &&g_{ij} (x)=\sum\limits_{k\in \cL_j \backslash \{i\}}
        (1+B_{jk} )^{-1}[ B_{jk} y_{jk} -A_{jk}
        \nonumber\\
        &&-x-v~\mbox{sgn}~(B_{jk} y_{jk} -A_{jk} -x)]
        \nonumber \\
        &&\times \Theta \left[ {\left| {B_{jk} y_{jk} -A_{jk} -x}
        \right|-v} \right]+\Lambda _i -y_{ij} .
\end{eqnarray}

The forward messages become
\begin{eqnarray}
\label{eq:AB_message_friction}
        A_{ij} &\leftarrow& -\mu _{ij} \ ,
        \\
        B_{ij} &\leftarrow& \frac{\Theta (-\mu _{ij} )}{\sum\limits_{k\in
        \cL_j \backslash \{i\}} {(1+B_{jk} )^{-1}\Theta \left[ {\left|
        {B_{jk} y_{jk} -A_{jk} -\mu _{ij} } \right|-v} \right]} },
\nonumber
\end{eqnarray}

To complete the algorithm,
we need information-provision messages
to determine the drawn current $y_{ij}$
at which the messages should be computed.
Analogous to the case of quadratic cost functions,
two methods are proposed.

In the method of backward information-provision messages,
the backward messages are computed directly
from the optimization of Eq.~(\ref{eq:energy_friction})
and sent from node $j$ to the descendent nodes, namely,
\begin{eqnarray}
\label{eq:backward_message_friction}
        &&y_{jk} \leftarrow (1+B_{jk})^{-1}
        [ B_{jk} y_{jk} -A_{jk} -\mu _{ij}
        -v~{\rm sgn}(B_{jk} y_{jk}
        \nonumber\\
        &&-A_{jk} -\mu _{ij} )]
        \Theta\left[ {\left| {B_{jk} y_{jk} -A_{jk} -\mu _{ij} } \right|-v}
        \right] \ .
\end{eqnarray}
This algorithm reduces the error at steady state to a level that
is still rather high. A careful examination of the solution shows
that the error is contributed by oscillatory solutions between
$y_{ij}$ and $y_{ji}$. Hence a learning rate $\eta $ is
introduced:
\begin{eqnarray}
        &&y_{jk}\leftarrow(1-\eta)y_{jk}+\eta(1+B_{jk})^{-1}[
        B_{jk}y_{jk}-A_{jk}
        \nonumber\\
        &&-\mu_{ij}-v~{\rm sgn}(B_{jk}y_{jk}-A_{jk}-\mu_{ij})]
        \nonumber\\
        &&\times\Theta\left[\left|
        B_{jk}y_{jk}-A_{jk}-\mu _{ij}\right|-v\right].
\label{eq:friction_learningrate}
\end{eqnarray}
The case $\eta= 1$ corresponds to the original algorithm.

In the method of forward information-provision messages,
a node first receives the messages from the ancestor
immediately before it updates its messages.
The working point is obtained by minimizing the energy
\begin{eqnarray}
        &&E_{ij}=A_{ij}\varepsilon_{ij}+\frac{1}{2}B_{ij}\varepsilon_{ij}^2
        +A_{ji}(-y_{ij}-\varepsilon_{ij}-y_{ji})
        \nonumber\\
        &&+\frac{1}{2}B_{ji}(-y_{ij}-\varepsilon_{ij}-y_{ji})^2
        +\frac{1}{2}(y_{ij}+\varepsilon_{ij})^2
        \nonumber\\
        &&+v\left|y_{ij}+\varepsilon_{ij}\right|,
\end{eqnarray}
with the optimal solution
\begin{eqnarray}
        &&y_{ij}\!\leftarrow\!(1+B_{ij}+B_{ji})^{-1}
        [B_{ij}y_{ij}-A_{ij}-B_{ji}y_{ji}+A_{ji}
        \nonumber\\
        &&-v~{\rm sgn}(B_{ij}y_{ij}-A_{ij}-B_{ji}y_{ji}+A_{ji})]
        \nonumber\\
        &&\times\Theta\left(\left|B_{ij}y_{ij}-A_{ij}-B_{ji}y_{ji}+A_{ji}
        \right|-v\right).
\end{eqnarray}
For further improvement, a learning rate is introduced, namely,
\begin{eqnarray}
        &&y_{ij}\!\leftarrow\!(1-\eta)y_{ij}
        +\eta(1+B_{ij}+B_{ji})^{-1}[B_{ij}y_{ij}-A_{ij}
        \nonumber\\
        &&-B_{ji}y_{ji}+A_{ji}
        -v~{\rm sgn}(B_{ij}y_{ij}-A_{ij}-B_{ji}y_{ji}+A_{ji})]
        \nonumber\\
        &&\times\Theta\left(\left|B_{ij}y_{ij}-A_{ij}-B_{ji}y_{ji}+A_{ji}
        \right|-v\right).
\label{ratefor}
\end{eqnarray}

\subsubsection{Simulation results}

\begin{figure}[t]
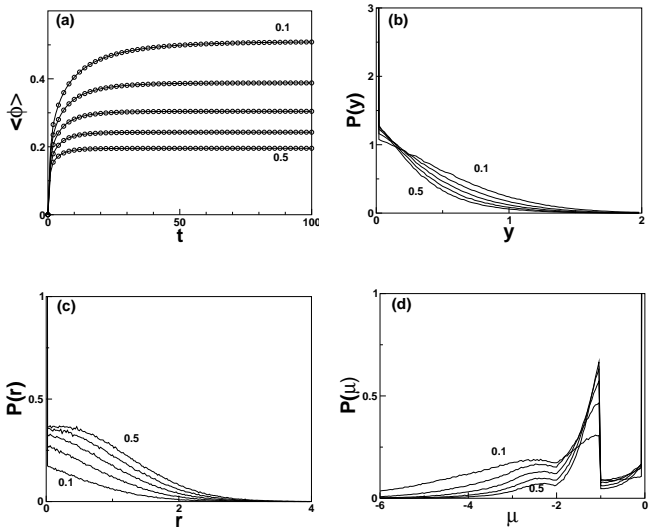

\centering
\begin{picture}(243,100)
\includegraphics[width=0.485\linewidth]{./fig6aM.eps}~~
\includegraphics[width=0.46\linewidth]{./fig6bM.eps}
\end{picture}
\vgap
\begin{picture}(243,100)
\includegraphics[width=0.475\linewidth]{./fig6cM.eps}~~
\includegraphics[width=0.475\linewidth]{./fig6dM.eps}
\end{picture}
\caption{Results for $N\!=\!1000$, $c\!=\!3$,
the frictional cost function with $v=1$ and 1000 samples.
(a) $\langle\phi\rangle$ obtained by the price iteration algorithm
as a function of $t$
for $\langle \Lambda\rangle\!=\! 0.1, 0.2, 0.3, 0.4, 0.5$
(top to bottom).
(b) The corresponding current distribution $P(y)$.
(c) The corresponding resource distribution $P(r)$.
(d) The corresponding chemical potential distribution $P(\mu)$.}
\label{fig:friction}
\end{figure}

To study the behavior of both price iteration and message passing
algorithms in the case of the frictional cost function,
we carried out simulations under
similar conditions to those of section~\ref{sec:numerical}.
Figure~\ref{fig:friction}(a) shows the average energy per link as
a function of iteration steps of the price iteration algorithm.
Figures~\ref{fig:friction}(b), (c) and (d)
show the current, resource, and chemical potential
distributions, $P(y)$, $P(r)$, and $P(\mu)$, respectively
for the various $\langle \Lambda \rangle $ values.

The results shown in Figs.~\ref{fig:friction}(a)-(c) exhibit a
similar qualitative behavior to those of the quadratic and
anharmonic cost functions.
However, there is a substantial difference
in the chemical potential distribution, shown in
Fig.~\ref{fig:friction}(d) as a pseudogap develops in the range
$v <\mu < 0$, as well as a kink at $\mu= -2v$. From
Eq.~(\ref{eq:friction_y}) one notices that a link becomes idle
when the potential difference at its vertices is less than $v$,
accounting for the existence of the pseudogap.

\begin{figure}[t]
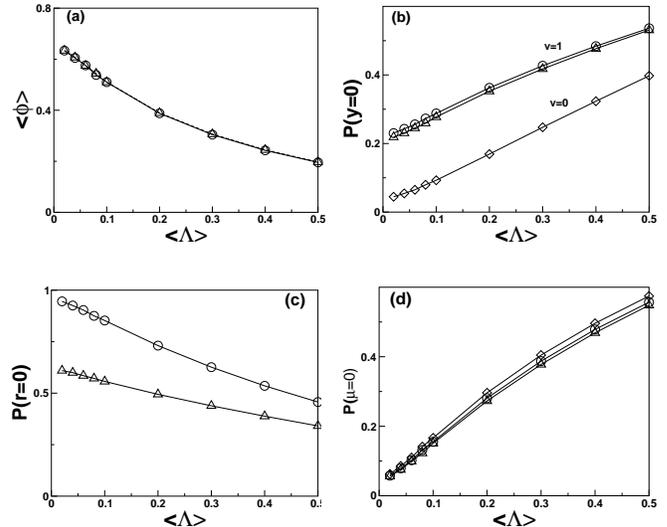

\centering
\begin{picture}(245,100)
\includegraphics[width=0.484\linewidth]{./fig7aM.eps}~
\includegraphics[width=0.484\linewidth]{./fig7bM.eps}
\end{picture}
\vgap
\begin{picture}(245,100)
\includegraphics[width=0.484\linewidth]{./fig7cM.eps}~
\includegraphics[width=0.484\linewidth]{./fig7dM.eps}
\end{picture}
\caption{Results for $N\!=\!1000$, $c\!=\!3$,
the frictional cost function with $v=1$, and 1000 samples.
(a) Average energy $\langle\phi\rangle$.
(b) The fraction of idle links $P(y=0)$.
(c) The fraction of unsaturated nodes $P(\mu=0)$.
(d) The fraction of saturated nodes $P(r=0)$.
Symbols in (a)-(d): results obtained
for the frictional cost function
by the price iteration algorithm ($\bigcirc$)
and the message-passing algorithm ($\square$);
results obtained for the quadratic cost function ($v=0$)
($\lozenge$).}
\label{fig:friction_comp}
\end{figure}

A quantitative comparison between the results obtained by price
iteration (\ref{eq:g_friction})
and message-passing~(\ref{eq:AB_message_friction})-
(\ref{eq:backward_message_friction})
algorithms ($\eta=1$, no learning rate) are presented in
Fig.~\ref{fig:friction_comp}. A comparison of the average energy
per link as a function of $\langle \Lambda \rangle $,
the fraction of idle links $P(y=0)$,
and the fraction of unsaturated nodes $P(\mu=0)$ are shown in
Figs.~\ref{fig:friction_comp}(a), (b) and (d), respectively,
showing good agreement
between the result obtained using both algorithms.
Results obtained by both price iteration and message passing
algorithms for a friction ($v=1$) cost are also contrasted with
results obtained for the quadratic ($v=0$) cost in
Figs.~\ref{fig:friction_comp}(b) and (d).

As shown in Fig.~\ref{fig:friction_comp}, the price iteration and
the original message-passing algorithms yield results agreeing in
the average energy (a), the fraction of idle links (b), and the
fraction of unsaturated nodes(d). Compared with the quadratic cost
function, the fraction of idle links is considerably increased
after introducing the friction, as shown in
Fig.~\ref{fig:friction_comp}(b). However, as shown in
Fig.~\ref{fig:friction_comp}(c), the message-passing algorithm
gives values much lower than those of price iteration, and is
inconsistent with the results in Fig.~\ref{fig:friction_comp}(d).

\begin{figure}[t]
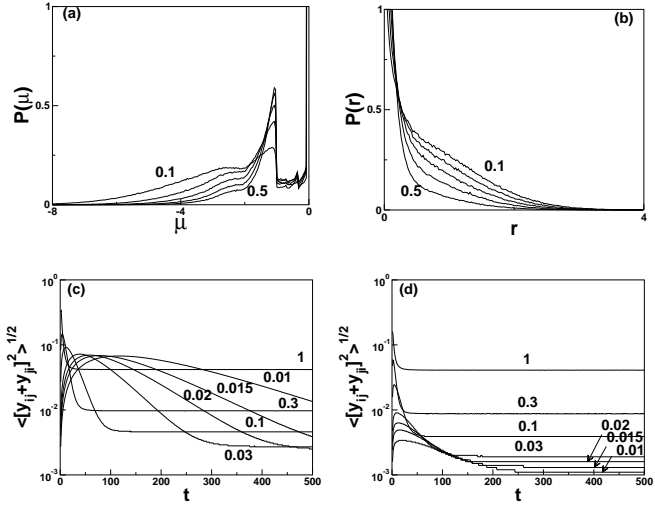

\centering
\begin{picture}(240,100)
\includegraphics[width=0.46\linewidth]{./fig8aM.eps}~~~
\includegraphics[width=0.465\linewidth]{./fig8bM.eps}
\end{picture}
\\
\begin{picture}(245,100)
\includegraphics[width=0.485\linewidth]{./fig8cM.eps}~
\includegraphics[width=0.485\linewidth]{./fig8dM.eps}
\end{picture}
\caption{Results for $N\!=\!1000$, $c\!=\!3$, the frictional cost
function with $v=1$, and 1000 samples. (a) The chemical potential
distribution $P(\mu)$ for $\langle \Lambda\rangle\!=\! 0.1, 0.2,
0.3, 0.4, 0.5$ (top to bottom). (b) The corresponding resource
distribution $P(r)$. (c)-(d) The convergence parameter $\langle
$[($y_{ij}$ + $y_{ji})$/2]$^{2}\rangle^{1/2}$ of the
message-passing algorithm as a function of iteration steps and for
various $\eta $ values, using backward and forward
information-provision messages in (c) and (d), respectively.}
\label{fig:friction_eta}
\end{figure}

The resource distribution in Fig.~\ref{fig:friction_eta}(a)
explains the discrepancy. Compared with the results in
Fig.~\ref{fig:friction}(c), the sharp peak at $r$ = 0 is broadened
to finite values of $r$. This shows that the original
message-passing algorithm is not precise in computing the
resources. Furthermore, the chemical potential distribution in
Fig.~\ref{fig:friction_eta}(b)
exhibits rough features in the pseudogap, and the
jumps near the edge of the pseudogap are less sharp than those in
Fig.~\ref{fig:friction}(d).


These unsatisfactory performances of the message-passing algorithm
can be traced to its non-convergence. In message-passing,
convergence is monitored by the root-mean-square average of
$\langle $[($y_{ij}$ + $y_{ji})$/2]$^{2}\rangle ^{1/2}$, which is
expected to approach 0. As shown in Fig.~\ref{fig:friction_eta}(c)
for the original algorithm ($\eta= 1$), the convergence parameter
reaches 0.04 at $t$ = 500, compared with the value of 0.0003 for
the price iteration algorithm.

To improve convergence, we introduce a learning rate according to
Eqs.~(\ref{eq:friction_learningrate}) and (\ref{ratefor}).
As shown in Figs.~\ref{fig:friction_eta}(c) and (d),
convergence improves for decreasing $\eta $,
but is also slowed down.
Comparing the two information-provision methods,
the one using forward information-provision messages
converges faster.

\begin{figure}[t]
\centerline{\hspace{-3mm}\epsfig{figure=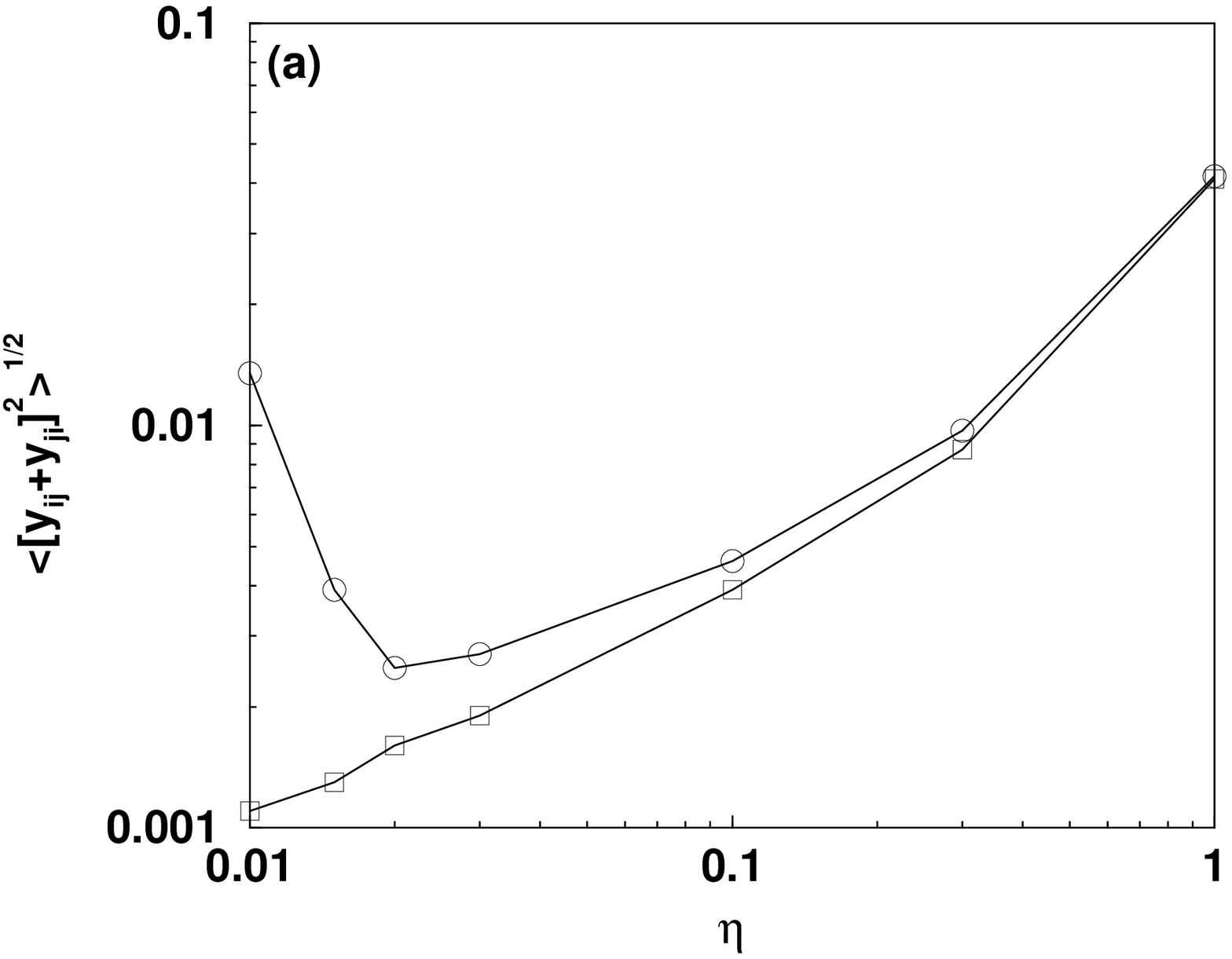,width=0.85\linewidth}}
\centerline{\epsfig{figure=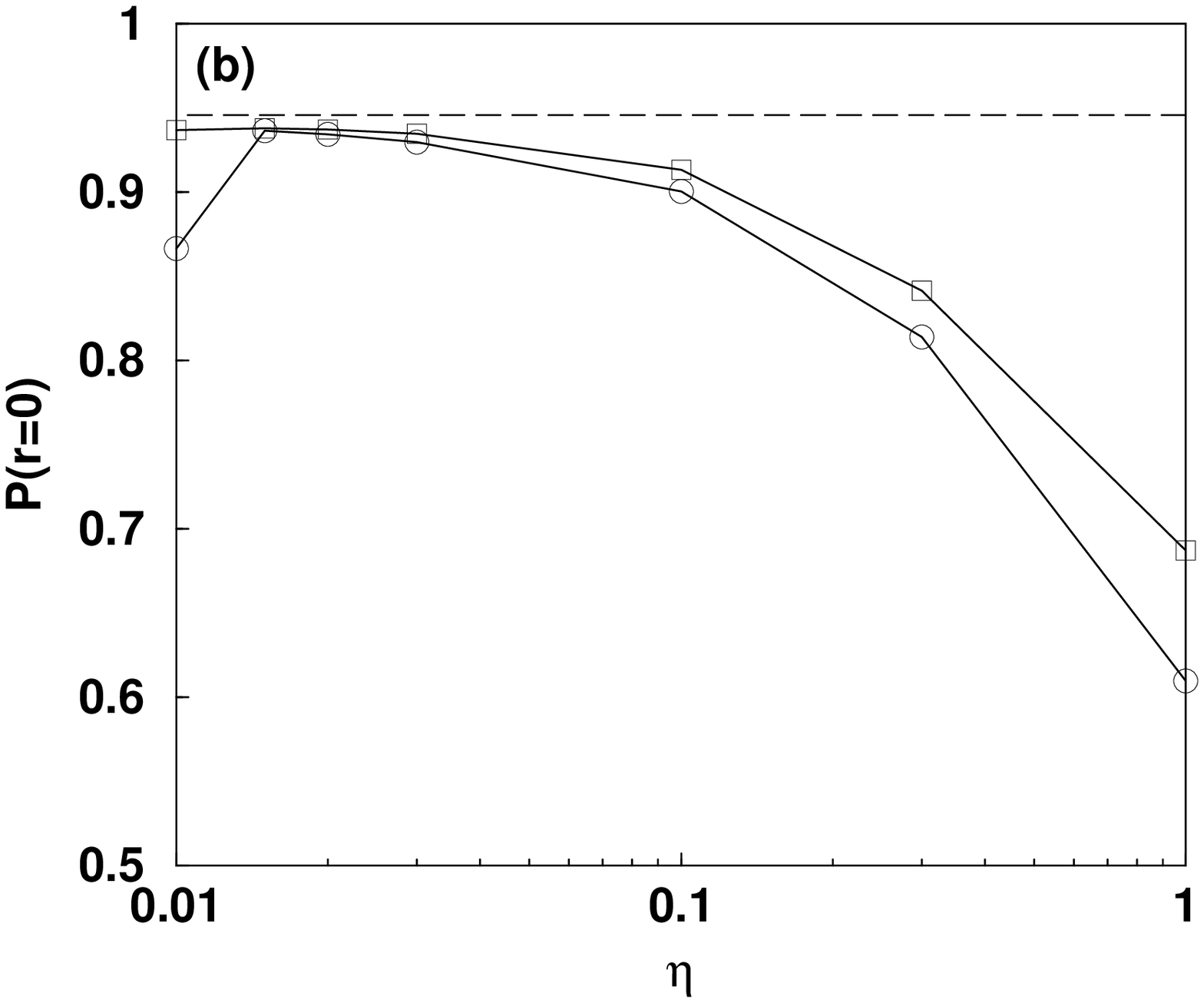,width=0.8\linewidth}}
\caption{The convergence parameter
$\langle $[($y_{ij}$ + $y_{ji})$/2]$^{2}\rangle^{1/2}$
of message-passing algorithms
as a function of the learning rate $\eta$
at $t=500$ with $N=1000$, $c=3$,
the frictional cost function with $v=1$, and 1000 samples.
Symbols: information-provision using backward messages ($\bigcirc$)
and forward messages ($\square$).
(b) The corresponding fraction of saturated nodes.
Symbols: same as in (a), and the price iteration algorithm (dashed line).}
\label{fig:friction_conv}
\end{figure}

As shown in Fig.~\ref{fig:friction_conv}(a),
better convergence is obtained
by the forward information-provision messages in 500 steps.
Fig.~\ref{fig:friction_conv}(b) summarizes the improvement in
 the fraction of saturated nodes on introducing the
learning rate for 500 steps; results obtained using the price iteration
algorithm are provided for comparison.
Obviously, further improvement can be made
by increasing the number of time steps,
and hence depends on the amount of computational resource
one wishes to commit.


\section{Conclusion}

The paper presents a study of inference and optimization tasks of
real value edges on sparse graphs under given constraints and cost
measure. A generic framework comprising of sparsely connected nodes,
representing constraints, and edges representing current variables
connecting them, is used as the basic framework for the inference
and optimization tasks. Inference of real values attributed to the
graph edges has very rarely been studied before within and outside
the statistical physics community. Although both theoretical
analysis and algorithmic solutions can be obtained for any
connectivity profile, we restricted this study to the case of fixed
connectivity - $c$.

The framework is analyzed using both the replica method and Bethe
approximation to obtain a set of recursive equations to be solved
numerically. The solutions provide numerical results for the free
energy, average energy, and the distribution of currents, resources,
and chemical potentials for the various cases.
The recursive equations also
enabled us to obtain scaling rules for various quantities of
interest as a function of the node connectivity. In addition, we
have devised message-passing and price iteration algorithms
for solving the optimization problem.
The message-passing algorithm is based on passing first and second
derivatives of the vertex free energy, representing the local
contribution to the system's free energy,
thus saving the need to pass
the full free energy functions as messages.
Despite the simplicity of the two-parameter messages, 
they yield exact solutions in the limit of sparse connectivity 
as long as they converge.

Most numerical studies have been carried out for the case of quadratic
cost that corresponds to the resource allocation problem which
initiated this study. In this case we fixed the nodes capacities,
representing biases in the local constraints, to quenched variables
drawn from some Gaussian distribution of given mean $\langle \Lambda
\rangle$ and unit variance. Numerical results for various parameters
values, $c$ and $\langle \Lambda \rangle$, show excellent agreement
between the analytical and algorithmic approaches both for finite and
asymptotic connectivity values.  Moreover, they expose an interesting
percolation transition of the clusters of nodes with negative
resources when $\langle\Lambda\rangle$ varies, giving rise to a
slowing down of the convergence of the saddle point equations below a
certain value of $\langle\Lambda\rangle$.

To study the efficacy of our approach to other cost measures we have
examined two different costs that include anharmonic and friction
terms.  We have applied two different algorithms in these cases based
on the price iteration and message-passing.  Price iterations involve
solving a nonlinear equation of the chemical potential at each step
numerically On the other hand, message-passing involves updating the
messages based on the working points estimated from the
information-provision messages. While the obtained solutions are
qualitatively similar to that of the quadratic cost, there are also
substantial algorithmic and conceptual differences, especially in the
case of friction cost. For the optimization task studied in this
paper, price iteration is simpler in implementation and converges
better when compared with message-passing.  However, for future
extensions to inference problems at finite temperatures, we expect the
message-passing approach to be more appropriate.  It is also useful to
adopt an adpative learning rate as a function of time to optimize the
performance~\cite{online_learning}.

We believe this research opens a rich area for further
investigations with many potential applications, especially when
additional restrictions are imposed and other costs considered.
More specifically, one may consider bandwidth limited links~\cite{iconip}
and other nonlinear costs which are of interest in realistic networks.
We expect that many nonlinear costs
may exhibit replica symmetry-breaking effects,
and it would be interesting to consider
how the analyses and algorithms should be modified
to cope with these effects.

\section*{Acknowledgements} This work is partially supported
by the Research Grant Council of Hong Kong
(grants HKUST6062/02P, DAG04/05.SC25 and DAG05/06.SC36)
and EVERGROW, IP No.~1935 in FP6 of the EU.

\appendix

\section{Replica Approach to Network Optimization}
\label{sec:app_replica}

To calculate the averaged replicated partition
function~(\ref{eq:replicatedpart}), we employ an integral
representation of the step function to obtain
\begin{eqnarray}
\label{eq:Z_replicated}
    &&\left\langle {Z^n} \right\rangle
    =\frac{1}{\cN }\sum\limits_{\cA_{ij} =0,1} \prod\limits_i \oint
    \frac{dz_i }{2\pi i z_i^{c+1}} \prod\limits_i
    z_i^{\sum_j \cA_{ij} }
    \nonumber\\
    &&\times\prod\limits_i \biggl[ \int d\Lambda _i
    \rho (\Lambda _i )\prod\limits_\alpha {\int {d\nu _i^\alpha }
    \int_{-\Lambda_i }^\infty {d\lambda _i^\alpha }
    \int \frac{d\hat {\lambda }_i^\alpha}{2\pi }}
    \nonumber \\
    &&\times e^{i\hat {\lambda }_i^\alpha \lambda _i^\alpha -\frac{\beta
    \epsilon }{2}(\nu _i^\alpha )^2} \biggr]
    \exp \biggl[ -\sum\limits_{i\alpha }
    {i\hat {\lambda}_i^\alpha }
    \sum\limits_j {\cA_{ij} (\nu _j^\alpha -\nu _i^\alpha)}
    \nonumber\\
    &&-\beta \sum\limits_{(ij)} \sum\limits_\alpha \cA_{ij} \phi
    (\nu _j^\alpha -\nu _i^\alpha ) \biggr]\ .
\end{eqnarray}
Collecting terms containing $\cA_{ij}$ and summing over them,
one obtains
\begin{eqnarray}
    &&\left\langle {Z^n} \right\rangle=\frac{1}{\cN}
    \prod\limits_i {\oint {\frac{dz_i }{2\pi iz_i^{c+1} }} }
    \prod\limits_i \Biggl[ \int d\Lambda _i
    \rho (\Lambda _i )
    \nonumber\\
    &&\times\prod\limits_\alpha {\int {d\nu _i^\alpha }
    \int_{-\Lambda_i }^\infty
    {d\lambda _i^\alpha } \int {\frac{d\hat {\lambda }_i^\alpha}{2\pi }}
    e^{i\hat {\lambda }_i^\alpha \lambda _i^\alpha -\frac{\beta
    \epsilon }{2}(\nu _i^\alpha )^2}} \Biggr]
    \nonumber \\
    &&\times \prod\limits_{(ij)} \Biggl[ 1+z_i z_j \exp \Biggl(
    \sum\limits_\alpha {(i\hat {\lambda }_i^\alpha
    -i\hat {\lambda }_j^\alpha)(\nu _i^\alpha -\nu _j^\alpha )}
    \nonumber\\
    &&-\beta \sum\limits_\alpha {\phi (\nu_j^\alpha -\nu_i^\alpha )}
    \Biggr) \Biggr] \ .
\end{eqnarray}
This includes a mixed term of $i$ and $j$ indices. An additional
expansion is required to disentangle the two indices.  The product over
(\textit{ij}) can be written as an exponential function whose argument
is
\begin{eqnarray}
\label{eq:ij_expansion}
    &&~\sum\limits_{ij}\sum\limits_{m=1}^\infty
    \frac{(-)^{m-1}}{2m}z_i^m z_j^m \exp \left( \sum\limits_\alpha
    im\hat\lambda_i^\alpha \nu _i^\alpha \right)
    \\
    &&\times\exp \left(
    \sum\limits_\alpha im\hat\lambda_j^\alpha \nu _j^\alpha\right)
    \sum\limits_{r_\alpha ,s_\alpha ,t_\alpha }
    \prod\limits_\alpha
    \frac{(-im\hat {\lambda }_i^\alpha \nu _j^\alpha )^{r_\alpha }}
    {r_\alpha!}
    \nonumber\\
    &&\times\frac{(-im\hat {\lambda }_j^\alpha \nu _i^\alpha )^{s_\alpha }}
    {s_\alpha!}
    \frac{(\nu _i^\alpha )^{t_\alpha }}{t_\alpha !}\left. {\left(
    {-\frac{d}{dy}} \right)^{t_\alpha }e^{-m\beta \phi (y)}}
    \right|_{y=\nu_j^\alpha },\nonumber
\end{eqnarray}
which gives rise to the mean-field parameters
\begin{eqnarray}
\label{eq:Q_rs}
    &&Q_{\mathbf{ {r}} ,\mathbf{ {s}} }^m
    =\frac{1}{\sqrt {cN} }\sum\limits_i z_i^m \exp \left(
    {\sum\limits_\alpha {im\hat {\lambda }_i^\alpha \nu _i^\alpha } }
    \right)
    \nonumber\\
    &&\times\prod\limits_\alpha {(-im\hat {\lambda }_i^\alpha
    )^{r_\alpha }(\nu _i^\alpha )^{s_\alpha }},
\end{eqnarray}
and the conjugate parameters $\hat {Q}_{\mathbf{ {r}} ,\mathbf{{s}} }^m $.
The replicated and averaged partition function $\langle Z^{n}\rangle$ becomes
\begin{eqnarray}
\label{eq:Zn_Q}
    &&\left\langle{Z^n} \right\rangle=\frac{1}{\cN}
    \prod\limits_{\mathbf{ {r}} ,\mathbf{ {s}} ,m} \int \frac{d\hat
    {Q}_{\mathbf{ {r}} ,\bar {s}}^m dQ_{\mathbf{ {r}} ,\bar {s}}^m
    }{2\pi /cN}
    \nonumber\\
    &&\times\exp \left( {-cN\sum\limits_{\mathbf{ {r}} ,\mathbf{
    {s}} ,m} {\hat {Q}_{\mathbf{ {r}} ,\bar {s}}^m Q_{\mathbf{ {r}}
    ,\bar {s}}^m } } \right)
    \prod\limits_i
    \Biggl[ \oint \frac{dz_i }{2\pi iz_i^{c+1} }\int d\Lambda _i
    \nonumber\\
    &&\times\rho(\Lambda_i)\prod\limits_\alpha {\left( {\int {d\nu
    _i^\alpha } \int_{-\Lambda _i }^\infty {d\lambda _i^\alpha } \int
    {\frac{d\hat {\lambda }_i^\alpha }{2\pi }} e^{i\hat {\lambda
    }_i^\alpha \lambda _i^\alpha -\frac{\beta
    \epsilon }{2}(\nu _i^\alpha )^2}} \right)} \Biggr]
        \nonumber \\
    &&\times \exp \Biggl\{ \sqrt {cN} \sum_{\mathbf{r},\mathbf{s},m}
    \hat Q_{\mathbf{r}, \mathbf{s}}^m
        \sum\limits_i z_i^m \exp \left( \sum\limits_\alpha
        im\hat\lambda_i^\alpha \nu _i^\alpha \right)
    \nonumber\\
    &&\times\prod\limits_\alpha (-im\hat\lambda_i^\alpha )^{r_\alpha}
        (\nu _i^\alpha )^{s_\alpha}
    +\sqrt {cN}\sum\limits_{\mathbf{r} ,\mathbf{s} ,m}
        \frac{(-)^{m-1}}{2m}
    \frac{Q_{\mathbf{r} ,\mathbf{s}}^m}
    {\prod\limits_\alpha {r_\alpha !s_\alpha !} }
    \nonumber\\
    &&\times\sum\limits_j
        z_j^m \exp \left( \sum\limits_\alpha im\hat\lambda_j^\alpha
        \nu _j^\alpha \right)
    \prod\limits_\alpha (\nu _j^\alpha )^{r_\alpha }
    \nonumber\\
    &&\times{\left(
        {-im\hat {\lambda }_j^\alpha -\frac{d}{dy}} \right)^{s_\alpha
        }e^{-m\beta \phi (y)}} \Biggr|_{y=\nu _j^\alpha } \Biggr\} .
\end{eqnarray}
The integration over $z_{i}$ is dominated by the term $m$ = 1 in
the $c^{th}$ order expansion of the exponential term that leads to
Eq.~(\ref{zn}). Both $Q_{{\mathbf {r}} ,{\mathbf {s}} } $ and
$\hat {Q}_{{\mathbf {r}} ,{\mathbf {s}} } $ are then given by the
saddle point equations
\begin{eqnarray}
\label{eq:saddlepoint}
        &&Q_{\mathbf{ {r}} ,\mathbf{ {s}} } =
        \frac{\sN_{1}}{\sD} \mbox{~~~~and~~~} \hat {Q}_{\mathbf{ {r}}
        ,\mathbf{ {s}} } = \frac{\sN_2}{\sD}
        \mbox{~~~~~~~where}
    \nonumber \\
        &&\sN_1 = \int {d\Lambda \rho (\Lambda )} \prod\limits_\alpha
        {\int {d\nu _\alpha } \int_{-\Lambda }^\infty {d\lambda _\alpha}
        \int {\frac{d\hat {\lambda }_\alpha }{2\pi }} }
    \nonumber\\
    &&\times\exp \left[
        {\sum\limits_\alpha {\left( {i\hat {\lambda }_\alpha (\lambda
        _\alpha +c\nu _\alpha )-\frac{\beta \epsilon }{2}(\nu_\alpha)^2
        } \right)} } \right]X^{{c-1}}
    \nonumber\\
    &&\times\prod\limits_\alpha {(-i\hat {\lambda }_\alpha )^{r_\alpha}
    (\nu_\alpha)^{s_\alpha } } \ ,
        \nonumber \\
        &&\sN_2 = \frac{1}{2}\int {d\Lambda \rho (\Lambda )}
        \prod\limits_\alpha {\int {d\nu _\alpha } \int_{-\Lambda }^\infty
        {d\lambda _\alpha } \int {\frac{d\hat {\lambda }_\alpha }{2\pi}}}
    \nonumber\\
    &&\times\exp \left[ {\sum\limits_\alpha
    {\left( {i\hat {\lambda }_\alpha
        (\lambda _\alpha +c\nu _\alpha )-\frac{\beta \epsilon }{2}
        (\nu_\alpha)^2 } \right)} } \right]X^{c-1}
    \nonumber \\
    &&\times\prod\limits_\alpha {\frac{(\nu _\alpha)^{r_\alpha }
        }{r_\alpha !}\frac{1}{s_\alpha !}\left. {\left( {-i\hat {\lambda
        }_\alpha -\frac{d}{dy}} \right)^{s_\alpha }e^{-\beta \phi (y)}}
        \right|} _{y=\nu _\alpha } ,
        \nonumber\\
        &&\sD = \int {d\Lambda \rho (\Lambda )} \prod\limits_\alpha {\int
        {d\nu _\alpha } \int_{-\Lambda }^\infty {d\lambda _\alpha } \int
        {\frac{d\hat {\lambda }_\alpha }{2\pi }} }
    \nonumber\\
    &&\times\exp \left[
        {\sum\limits_\alpha {\left( {i\hat {\lambda }_\alpha (\lambda
        _\alpha +c\nu _\alpha )-\frac{\beta \epsilon }{2}(\nu _\alpha)^2
        } \right)} } \right]X^c \ ,
\end{eqnarray}
where $X$ is given by Eq.~(\ref{xx}).
By virtue of the saddle point equations (\ref{eq:saddlepoint}),
one can show that
\begin{eqnarray}
\label{eq:QQhat}
        &&\hat Q_{\mathbf{r} ,\mathbf{s} }
        =\frac{1}{2}\sum\limits_{\mathbf{t} ,\mathbf{u} }
        \prod\limits_\alpha \frac{(-)^{t_\alpha }}{r_\alpha !t_\alpha
        !(s_\alpha -t_\alpha )!u_\alpha !}
    \nonumber\\
    &&\times\left( \frac{d}{dy}\right)^{t_\alpha +u_\alpha }
    e^{-\beta \phi (y)} \Biggr|_{y=0}
        Q_{\mathbf{s} -\mathbf{t} ,\mathbf{r} +\mathbf{u} }\ .
\end{eqnarray}
Exploiting the even nature of $\phi (y)$ and relation
(\ref{eq:QQhat})~\cite{footnote1}, 
the expressions for $X$ and $Q_{\mathbf{r},\mathbf {s}}$ reduce to
\begin{equation}
\label{eq:X_expression}
        X\!=\!\sum\limits_{\mathbf{r},\mathbf{s}}
        \frac{Q_{\mathbf{r} ,\mathbf{s}}}
        {\prod\limits_\alpha{r_\alpha !s_\alpha !} }
        \prod\limits_\alpha {(\nu _\alpha)^{r_\alpha }
        \!\!\left.{\left( {-i\hat {\lambda }_\alpha
        \!-\!\frac{d}{dy}} \right)^{s_\alpha }e^{-\beta \phi (y)}}
        \right|_{y=\nu _\alpha } }.
\end{equation}

To better understand the symmetry properties of the order parameters,
we consider the generating function $P_{\mathbf{ {s}} } (\mathbf{ {z}} )$
and its inversion in Eq.~(\ref{genf}).
Substituting Eq.~(\ref{eq:X_expression}) into Eq.~(\ref{genf}),
we reproduce Eq.~(\ref{eq:P_sDef}), with $\sD_P$ being
\begin{eqnarray}
        &&\sD_P
        = \int {d\Lambda \rho (\Lambda )} \prod\limits_\alpha \left[
        \int {d\nu _\alpha } \int_{-\Lambda }^\infty {d\lambda _\alpha }
        \int \frac{d\hat {\lambda }_\alpha }{2\pi }\right.
    \nonumber\\
    &&\left.\times\exp \left( i\hat {\lambda}_\alpha
        (\lambda _\alpha +c\nu _\alpha)
    -\frac{\beta \epsilon}{2}(\nu _\alpha)^2 \right) \right]
    \sum\limits_{\mathbf{s}_k} \prod\limits_{k=1}^c
    \nonumber \\
        &&\times P_{\mathbf{s}_k}(\mathbf{\nu})
        \prod\limits_{k\alpha } \frac{1}{s_k^\alpha !}
        \left.\left(-i\hat\lambda_\alpha -\frac{d}{dy}\right)^{s_k^\alpha }
    e^{-\beta \phi(y)} \right|_{y=\nu_\alpha}.
    \nonumber\\
\label{eq:Dpdef}
\end{eqnarray}

Once we have represented the order parameters $Q_{\mathbf{ {r}}
,\mathbf {s}}$ using the generating function $P_{\mathbf{ {s}} }
(\mathbf{ {z}} )$, we can make explicit assumptions about their
symmetry properties. In particular, in the replica symmetric ansatz,
we consider functions of the form Eq.~(\ref{moments}).

Notice that the replicas in Eq.~(\ref{moments}) are coupled through
their common dependence on the disordered distribution of $\Lambda
$. This is different from the SK model, in which the dependence on
the disorder is integrated out, and the interaction between the
replicas is explicit.
Using the ansatz (\ref{moments}), the recursion relation for
$P_{\mathbf{ {s}} } (\mathbf{ {z}} )$ can be replaced by a
recursion relation for the function $R$ in Eq.~(\ref{eq:R_T}),
where
\begin{eqnarray}
\label{eq:R_TD}
        &&\sD_R = \left\langle \left\{ \int d\nu
        \prod\limits_{k=1}^c {\left[ {\int {d\nu _k
        R(\nu ,\nu _k \vert {\rm {\bf T}}_k )} } \right]}\right.\right.
        \nonumber\\
        &&\left.\left.\times
        \Theta \left( {\sum\limits_{k=1}^c {\nu _k } -c\nu+\Lambda _V}
        \right) \right. \right.
        \\
        &&\times\left. \left.
        \exp \left[ {-\frac{\beta \epsilon }{2}\nu ^2
        -\beta \sum\limits_{k=1}^c
        {\phi \left( {\nu -\nu _k } \right)} } \right] \right\}^n
        \right\rangle_\Lambda^{\frac{1}{n}} \ .\nonumber
\end{eqnarray}
$\Lambda _{V}$ is the capacity of the vertex fed by $c$ trees
\textbf{T}$_{1}$, {\ldots}, \textbf{T}$_{c}$.

Letting $y\equiv\nu-z$, 
we consider solutions of Eq.~(\ref{eq:R_T}) in the form
\begin{equation}
\label{eq:R_W}
        R(z,\nu \vert {\rm {\bf T}})=W(\nu)Z_V (y\vert{\rm {\bf T}}).
\end{equation}
Separating the dependence on the current potentials from that on
the currents, the extra Gaussian distribution of $\nu $ in
Eq.~(\ref{eq:R_TD}) prevents the integration of $\nu $ from
diverging. Indeed, in the $n\to 0$ limit and as $\epsilon\to 0$, 
the function  $W(z)$ becomes independent of $z$ and can be represented 
as
\begin{equation}
\label{eq:W}
        W(\nu)=\sqrt[4]{\frac{\beta\epsilon}{2\pi}}.
\end{equation}
The recursion relation involving the currents becomes decoupled to give
\begin{eqnarray}
        &&Z_V(y\vert {\mathbf T})
        =\prod\limits_{k=1}^{c-1} \left[
        \int {dy_k Z_V (y_k \vert {\mathbf T}_k )} \right]
        \nonumber\\
    &&\times\Theta \left( \sum\limits_{k=1}^{c-1} {y_k}-y\right.
        +\Lambda _{V({\mathbf T})}\Biggr)
    \exp\left[-\beta\sum\limits_{k=1}^{c-1}{\phi(y_k)}\right]
    \nonumber\\
    &&\times\exp\left\{-\left\langle \ln \left\{ \prod\limits_{k=1}^c
        {\left[ {\int {dy_k Z_V (y_k \vert {\mathbf T}}_k ) } \right]}
    \right.\right.\right.
        \\
    &&\left.\left.\left.\times
    \Theta \left( {\sum\limits_{k=1}^c {y_k } +\Lambda _V }
        \right)\exp \left[ {-\beta \sum\limits_{k=1}^c {\phi (y_k )}}
        \right] \right\} \right\rangle _\Lambda\right\} \ .
    \nonumber
\end{eqnarray}
Let $F_{V}(y\vert $\textbf{T}) be the vertex free energy when a
current $y$ is drawn from the vertex of a tree \textbf{T}, given by
$F_V (y\vert {\rm {\bf T}})=-T\ln Z_V (y\vert {\rm {\bf T}})$.
Then the recursion relation of the free energy is given by
Eq.~(\ref{recurt}), which in the zero-temperature limit becomes
Eq.~(\ref{recur}).

To calculate the free energy in the replica approach,
one returns to Eq.~(\ref{zn}).
In the second term of the exponential argument therein,
one eliminates $\hat {Q}_{\mathbf{ {r}} ,\mathbf{ {s}} }$
by Eq.~(\ref{eq:QQhat}),
expresses $Q_{\mathbf{ {r}} ,\mathbf{ {s}} } $ in terms of
$P_{\mathbf{ {s}}} (\mathbf{ {z}} )$ by Eq.~(\ref{genf}) and,
in turn, $R(z_\alpha,\nu\vert \Lambda )$ by Eq.~(\ref{moments}).
In the third term, one expresses $X$ in terms of
$Q_{\mathbf{ {r}} ,\mathbf{ {s}} } $ by Eq.~(\ref{eq:X_expression})
and follow similar steps.
The result is
\begin{eqnarray}
        &&\left\langle {Z^n} \right\rangle \!=\!\exp N\left\{
        \frac{c}{2}\!-\!\frac{c}{2}\left\langle \left\{\! \int {d\nu _1
        } {d\nu _2 } R(\nu_2, \nu_1 \vert {\rm {\bf T}}_1 )
    \right.\right.\right.
        \nonumber\\
    &&\times R(\nu_1, \nu_2 \vert {\mathbf T}_2 )
    \exp \left[ {\!-\!\beta \phi(\nu_1\!-\!\nu_2 )} \right]
    \biggr\}^n \biggr\rangle_\Lambda
    \\
        &&+\ln \left\langle \left\{
        \int\!{d\nu } \prod\limits_{k=1}^c {\left[
        {\int \!{d\nu_k } R(\nu, \nu_k \vert {\rm {\bf T}}_k )} \right]}
    \Theta \!\left( \sum\limits_{k=1}^c {\nu _k }
        \right.\right.\right.
    \nonumber\\
    &&\left.\left.
        \!-\!c\nu \!\!+\Lambda \Biggr)
        \exp \left( {\!-\!\beta \sum\limits_{k=1}^c {\phi (\nu
        \!-\!\nu_k )} -\frac{\beta\epsilon}{2}\nu^2
        } \right) \right\}^n \right\rangle _\Lambda.
    \nonumber
\end{eqnarray}
Using the recursion relation Eq.~(\ref{eq:R_T}),
one can show that the sum of the first two terms
in the exponential argument vanishes.
In the limit $n\rightarrow 0$ one obtains the free energy
\begin{eqnarray}
        &&\left\langle {\beta F} \right\rangle =
        -N\left\langle \ln \left\{
        \int {d\nu } \prod_{k=1}^c \left[
        R(\nu, \nu _k \vert {\mathbf T}_k) \right]
    \Theta \left( \sum_{k=1}^c\nu_k\right.\right.\right.
        \nonumber\\
    &&\left.\left.- c \nu + \Lambda \Biggr)
        \exp \left[ {-\beta \sum_{k=1}^c {
        \phi (\nu -\nu _k )} -\frac{\beta\epsilon}{2} \nu^2 }\right]
        \right\} \right\rangle _\Lambda .
    \nonumber\\
\label{eq:freeR}
\end{eqnarray}
Using the vertex free energy representation,
one then straightforwardly rewrites
Eq.~(\ref{eq:freeR}) as Eq.~(\ref{free}) (up to a constant).

\section{Messages in the Bayesian approximation}
\label{sec:app_Bayesnets}

To show that the vertex free energies are directly related to passed
messages in the Bayesian approximation,
one resorts to formulating the problem on a bipartite
graph and deriving the closed set of equations that relate to the
messages passed from variables to interaction nodes and vice versa.

\begin{figure}[t]
\centerline{\epsfig{figure=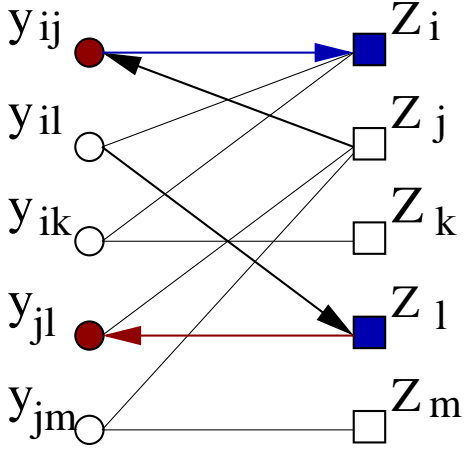,width=0.7\linewidth}}
\caption{A bipartite graph representation of the resource
allocation problem with the current variables
 $y$ on the left and the interaction variables $Z$ on the right.}
\label{fig:bipartite}
\end{figure}

The representation of the problem as a bipartite graph is shown
in Fig.~\ref{fig:bipartite}, with the current variables
$y$ on the left and the interaction variables $Z$ on the right.
Using conventional notations~\cite{mackaybook} one can easily
derive the closed set of equations:
\begin{equation}
        Q(y_{ij})\propto\!P(y_{ij})~R(y_{ij})
\label{eq:Q}
\end{equation}
where $Q(y_{ij})$ is the posterior of $y_{ij}$ given $Z_j$,
$P(y_{ij})$ is the prior of $y_{ij}$,
and $R(y_{ij})$ is the likelihood of $Z_j$ given $y_{ij}$.
As shown in Fig.~\ref{fig:bipartite},
the message from $Z_j$ to $y_{ij}$ is $Q(y_{ij})$,
and the message from $y_{ij}$ to $Z_i$ is $R(y_{ij})$. Thus,
\begin{eqnarray}
        &&R(y_{ij})=\int\!\!\prod_{k\in\cL_i)\setminus j}
        \!\!dy_{jk} ~P(Z_j\mid y_{ij},\{y_{jk}:k\in\cL_j \setminus
        i\} )
    \nonumber\\
    &&\times\prod_{k\in\cL_j \setminus i} \!\!Q(y_{jk}) \ .
\label{eq:P}
\end{eqnarray}
Using
\begin{eqnarray}
        &&P(y_{ij})\propto \exp\left(-\frac{\beta}{2} \phi(y_{ij})\right)
        \\
        && P(Z_j\mid y_{ij},\{y_{jk}:k\in\cL_j \setminus i\} )
        \nonumber\\
    &&\propto \Theta \left(\Lambda_{j}- y_{ij} + \sum_{k\in \cL_j
       \setminus i} y_{jk} \right) \ ,
\end{eqnarray}
and substituting the expression for $Q$ (\ref{eq:Q}) into the $P$
equation (\ref{eq:P}) one obtains
\begin{eqnarray}
        &&R(y_{ij})\!\propto\!\prod_{k\in\cL_j \setminus i}
        \!\left(\int dy_{jk} \right)\!\Theta\!\left(\Lambda_{j}- y_{ij} +
        \sum_{k\in\cL_j \setminus i} y_{jk}\right)
        \nonumber\\
    &&\times\prod_{k\in\cL_j \setminus i}
        \exp\left(-\frac{\beta}{2} \phi(y_{jk})\right)~R(y_{jk}) \ .
\label{rij}
\end{eqnarray}
Let $\tilde F_V(y_{ij}|Z_j) = -T\ln R(y_{ij})$.
Then on taking the logarithm of both sides of Eq.~(\ref{rij})
and normalizing,
one retrieves Eq.~(\ref{free})
if $\tilde F_V(y_{ij}|Z_j)$ is identified
with the vertex free energy $F_V(y_{ij}|{\mathbf T}_j)$,
\begin{eqnarray}
        &&\tilde F_V(y_{ij}|Z_j)=-T\ln\Biggl\{
        \prod_{k=1}^{c-1}\left(\int dy_{jk}\right)
        \Theta\left(\sum_{k=1}^{c-1}y_{jk}-y_{ij}\right.
    \nonumber\\
    &&+\Lambda_j\Biggr)
        \exp\left[-\beta\sum_{k=1}^{c-1}
        \left(\tilde F_V(y_{jk}|Z_k)
        +\phi(y_{jk})\right)\right]\Biggr\}
        -F_{\rm av}.
    \nonumber\\
\end{eqnarray}
This means that the vertex free energy $F_V(y_{ij}|{\mathbf T}_j)$
is equivalent to $-T$ times the logarithm of the message $R(y_{ij})$
from $y_{ij}$ to $Z_i$.



\begin{thebibliography}{0}

\bibitem{Nishimori_book} H.~Nishimori, {\it Statistical Physics of Spin
Glasses and Information Processing}, (Oxford University Press,
Oxford, UK, 2001).

\bibitem{os} M.~Opper and D.~Saad, {\it Advanced Mean Field Methods}
(MIT Press, Cambridge, MA, 2001).

\bibitem{mackaybook} D.~J.~C.~Mackay,
{\it Information Theory, Inference and Learning Algorithms},
(Cambridge University Press, Cambridge, UK, 2003).

\bibitem{yedidia} J.~S.~Yedidia, W.~T.~Freeman and Y.~Weiss,
IEEE Trans. IT {\bf 51} 2282 (2005).

\bibitem{mezard}
M.~M\'ezard, arXiv:cond-mat/0401237 (2004).

\bibitem{MZreview} M.~M\'{e}zard and R.~Zecchina, Phys.~Rev.~E {\bf
66}, 056126 (2002)

\bibitem{KSreview} Y.~Kabashima and D.~Saad, J.~Phys.~A {\bf 37},
R1 (2004).

\bibitem{saul94} L.~Saul and M.~Jordan, Neural Computation {\bf 6},
1174 (1994).

\bibitem{Lauritzen}S. L.~Lauritzen, J.~of the American Stat.~Assoc.
{\bf 87}, 1098 (1992).

\bibitem{skantzos} N.~Skantzos, I.~P.~Castillo, and J.~P.~L.~Hatchett,
Phys.~Rev.~E {\bf 72}, 066127 (2005).

\bibitem{us_nips} K.~Y.~M.~Wong, D.~Saad and Z.~Gao,  {\it Advances in
Neural Information Processing Systems} {\bf 18},
Y.~Weiss, B.~Sch\"{o}lkopf, and J.~Platt (eds.)
(MIT Press, Cambridge, MA, 2005) 1529.

\bibitem{us_prerc} K.~Y.~M.~Wong and D.~Saad, Phys.~Rev.~E {\bf 74}, 010104(R)
(2006).

\bibitem{bertsekas} D.~Bertsekas, Linear Network Optimization (MIT
Press, Cambridge, MA, 1991).

\bibitem{resourceallocation} L.~Peterson and B.~S.~Davie, {\it Computer
Networks: A Systems Approach}, (Academic Press, San Diego CA, 2000).

\bibitem{resourceallocation2}
Y.~C.~Ho, L.~Servi, and R.~Suri, Large Scale Syst. {\bf 1}, 51 (1980).

\bibitem{Shenker}
S.~Shenker, D.~Clark, D.~Estrin and S.~Herzog,
ACM Comp.~Comm.~Rev. {\bf 26}. 19 (1996).

\bibitem{om} R.~L.~Rardin, {\it Optimization in Operations Research}
(Prentice Hall, NJ, 1998)

\bibitem{wong87}
K.~Y.~M.~Wong and D.~Sherrington, J.~Phys.~A {\bf 20}, L793 (1987).

\bibitem{footnote}
This term is marginalized over all inputs to the current vertex, leaving
the difference in current potentials $y$ as its sole argument, hence
the terminology used.

\bibitem{footnote1}
If $\phi(y)$ is not an even function of $y$, 
analyses along the lines of Appendix A show that 
$\phi(y_k)$ in the Boltzmann factors of Eq.~(\ref{recurt}) 
have to be replaced by $\phi(\sigma_k y_k)$, 
where $\sigma_k=\pm 1$ should be quench averaged.

\bibitem{footnote2}
Compared with the results in Fig. 2(a) inset of \cite{us_prerc}, 
here we have an even faster convergence to the high connectivity limit, 
after we have separated the contribution of the continuous component 
at $y=0$.

\bibitem{kelly}
F.~P.~Kelly, Euro. Trans. on Telecommunications \& Related Technologies
{\bf 8}, 33 (1997).

\bibitem{iconip}
K.~Y.~M.~Wong, C.~H.~Yeung, and D.~Saad,
{\it Lecture Notes in Computer Science} {\bf 4233}, Part II,
I. King {\it et al} (eds.), 754 (Springer-Verlag, Berlin, 2006).

\bibitem{online_learning}
{\it On-line Learning in Neural Networks}, D.~Saad (ed.)
(Cambridge University Press, Cambridge, UK, 1998).

\end{thebibliography}
\end{document}